\documentclass{aa}                   
\usepackage{graphicx}                
\usepackage{txfonts}                 

\newcommand{\kms}{km\,s$^{-1}$}
\newcommand{\ms}{m\,s$^{-1}$}

\begin{document}

\title{High-resolution spectroscopy and spectropolarimetry of the total lunar eclipse January 2019\thanks{Based on data acquired with the Potsdam Echelle Polarimetric and Spectroscopic Instrument (PEPSI) using the Large Binocular Telescope (LBT) in Arizona.
}}

\author{K. G. Strassmeier\inst{1,2}, I. Ilyin\inst{1}, E. Keles\inst{1}, M. Mallonn\inst{1}, A. J\"arvinen\inst{1}, M. Weber\inst{1}, F. Mackebrandt\inst{3}, \and J. M. Hill\inst{4}}

\institute{
    Leibniz-Institute for Astrophysics Potsdam (AIP), An der Sternwarte 16, D-14482 Potsdam, Germany; \\ \email{kstrassmeier@aip.de},
    \and
    Institute for Physics and Astronomy, University of Potsdam, Karl-Liebknecht-Str. 24/25, D-14476 Potsdam, Germany;
    \and
    Max-Planck-Institut f\"ur Sonnensystemforschung, Justus-von-Liebig-Weg 3, D-37077 G\"ottingen, Germany; Institute for Astrophysics, University of G\"ottingen, Germany;
    \and
    Large Binocular Telescope Observatory (LBTO), 933 N.\,Cherry Ave., Tucson, AZ 85721, U.S.A.}

\date{Received ... ; accepted ...}

\abstract{Observations of the Earthshine off the Moon allow for the unique opportunity to measure the large-scale Earth atmosphere. Another opportunity is realized during a total lunar eclipse which, if seen from the Moon, is like a transit of the Earth in front of the Sun.}{We thus aim at transmission spectroscopy of an Earth transit by tracing the solar spectrum during the total lunar eclipse of January 21, 2019.}{Time series spectra of the Tycho crater were taken with the Potsdam Echelle Polarimetric and Spectroscopic Instrument (PEPSI) at the Large Binocular Telescope (LBT) in its polarimetric mode in Stokes IQUV at a spectral resolution of 130\,000 (0.06\,\AA ). In particular, the spectra cover the red parts of the optical spectrum between 7419--9067\,\AA . The spectrograph's exposure meter was used to obtain a light curve of the lunar eclipse.}{The brightness of the Moon dimmed by 10\fm75 during umbral eclipse. We found both branches of the O$_2$ A-band almost completely saturated as well as a strong increase of H$_2$O absorption during totality. A pseudo O$_2$ emission feature remained at a wavelength of 7618\,\AA , but it is actually only a residual from different P-branch and R-branch absorptions. It nevertheless traces the eclipse. The deep penumbral spectra show significant excess absorption from the Na\,{\sc i} 5890-\AA\ doublet, the Ca\,{\sc ii} infrared triplet around 8600\,\AA, and the K\,{\sc i} line at 7699\,\AA\ in addition to several hyper-fine-structure lines of Mn\,{\sc i} and even from Ba\,{\sc ii}. The detections of the latter two elements are likely due to an untypical solar center-to-limb effect rather than Earth's atmosphere. The absorption in Ca\,{\sc ii} and K\,{\sc i} remained visible throughout umbral eclipse. Our radial velocities trace a wavelength dependent Rossiter-McLaughlin effect of the Earth eclipsing the Sun as seen from the Tycho crater and thereby confirm earlier observations. A small continuum polarization of the O$_2$ A-band of 0.12\,\% during umbral eclipse was detected at 6.3$\sigma$. No line polarization of the O$_2$ A-band, or any other spectral-line feature, is detected outside nor inside eclipse. It places an upper limit of $\approx$0.2\% on the degree of line polarization during transmission through Earth's atmosphere and magnetosphere.}{}


\authorrunning{Strassmeier et al.}

\titlerunning{High-resolution spectroscopy of the lunar eclipse Jan. 2019}

\maketitle

\section{Introduction}

Biological activity on Earth has many by-products, such as oxygen and ozone in association with water vapor, methane, and carbon dioxide (Lovelock \cite{love}, Des Marais et al. \cite{desm}). These biogenic molecules present attractive narrow molecular bands at optical and near infrared wavelengths for the detection in atmospheres of other planets. Our own Earth provides the unique opportunity to investigate these features, either through Earthshine observations or through lunar eclipses. Earthshine is the sunlight reflected off the Earth that illuminates the otherwise unlit side of the Moon and back reflects onto Earth where we can see it (for reviews see Goode et al. \cite{goode01} or Thejll et al. \cite{thejll}). In taking the Earth as the prototype of a habitable planet, Earthshine observations provide the possibility to verify the biogenic presence with the same techniques that otherwise are being used for observing stars. Arnold et al. (\cite{luc}) and Seager et al. (\cite{seager}) used Earthshine observations to prove that the spectral signature of photosynthetic pigments and atmospheric biogenic molecules are indeed detectable, suggesting that life on other planets could be detected in a similar way. Spectroscopic and spectro-polarimetric studies in general (Arnold et al. \cite{luc}, Woolf et al. \cite{woolf}, Montan\'es-Rodriguez et al. \cite{mont04}, Seager et al. \cite{seager}, Montan\'es-Rodriguez et al. \cite{mont}, Hamdani et al. \cite{ham}, Pall\'e et al.~\cite{palle}, Sterzik et al. \cite{sterz}, Miles-P\'aez et al. \cite{mil:pal}) have revealed, for example, such properties as the variability of the vegetation red edge as a function of Earth's cloud-cover.

The most prominent optical molecular feature identified in the Earthshine spectrum is the O$_2$ A-band at 760\,nm (e.g., Fauchez et al. \cite{fau} and references therein). Others are the B-band at 690\,nm and bands from water vapor in the intervals 653--725\,nm and 780--825\,nm as well as in the near infrared at 930\,nm and 1.12\,$\mu$m. Ozone in the optical is comparably only a weak absorber, but it is present over a long wavelength range known as the Chappuis band (500--700\,nm). However, ozone is the dominant absorber in the ultraviolet. Stam (\cite{stam}) had already predicted several of these bands from radiative transfer simulations of reflective light from Earth-like atmospheres and codes are now publicly available that can remove telluric contamination from stellar spectra to a high precision (e.g., Smette et al. \cite{molecfit}). Aside from molecular features, optical atomic lines in Earthshine or eclipse spectra that have been detected so far from high-resolution spectroscopy are only the sodium D lines (Vidal-Madjar et al.~\cite{vidal}). Other metal lines (Fe, Li, K, Mg, Ca) were unconfirmed from high-resolution observations (Gonz\'alez-Merino et al. \cite{sarg}), despite earlier claims of detections of Ca\,{\sc ii} H\&K and the Ca\,{\sc ii} infrared triplet (IRT) at low resolution by Pall\'e et al.~(\cite{palle}) (see the discussions in Vidal-Madjar et al.~\cite{vidal} and Garc\'ia Munoz et al.~\cite{garcia}).

\begin{figure*}
{\bf a.} \hspace{90mm} {\bf b.}\\
\includegraphics[angle=0,width=80mm,clip]{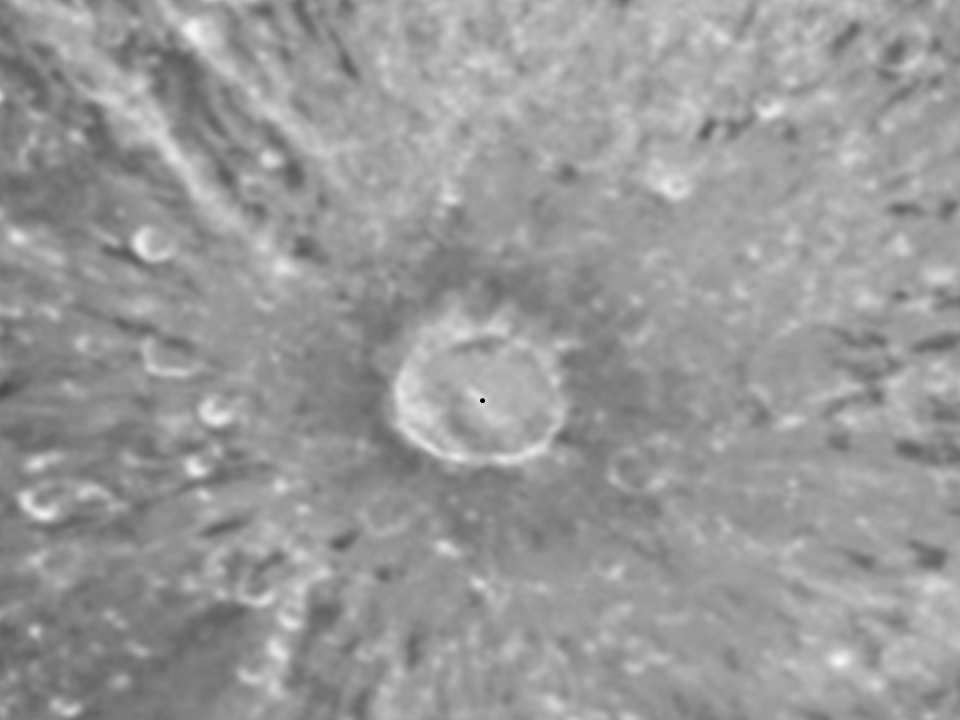}\hspace{5mm}
\includegraphics[angle=0,width=88mm,clip]{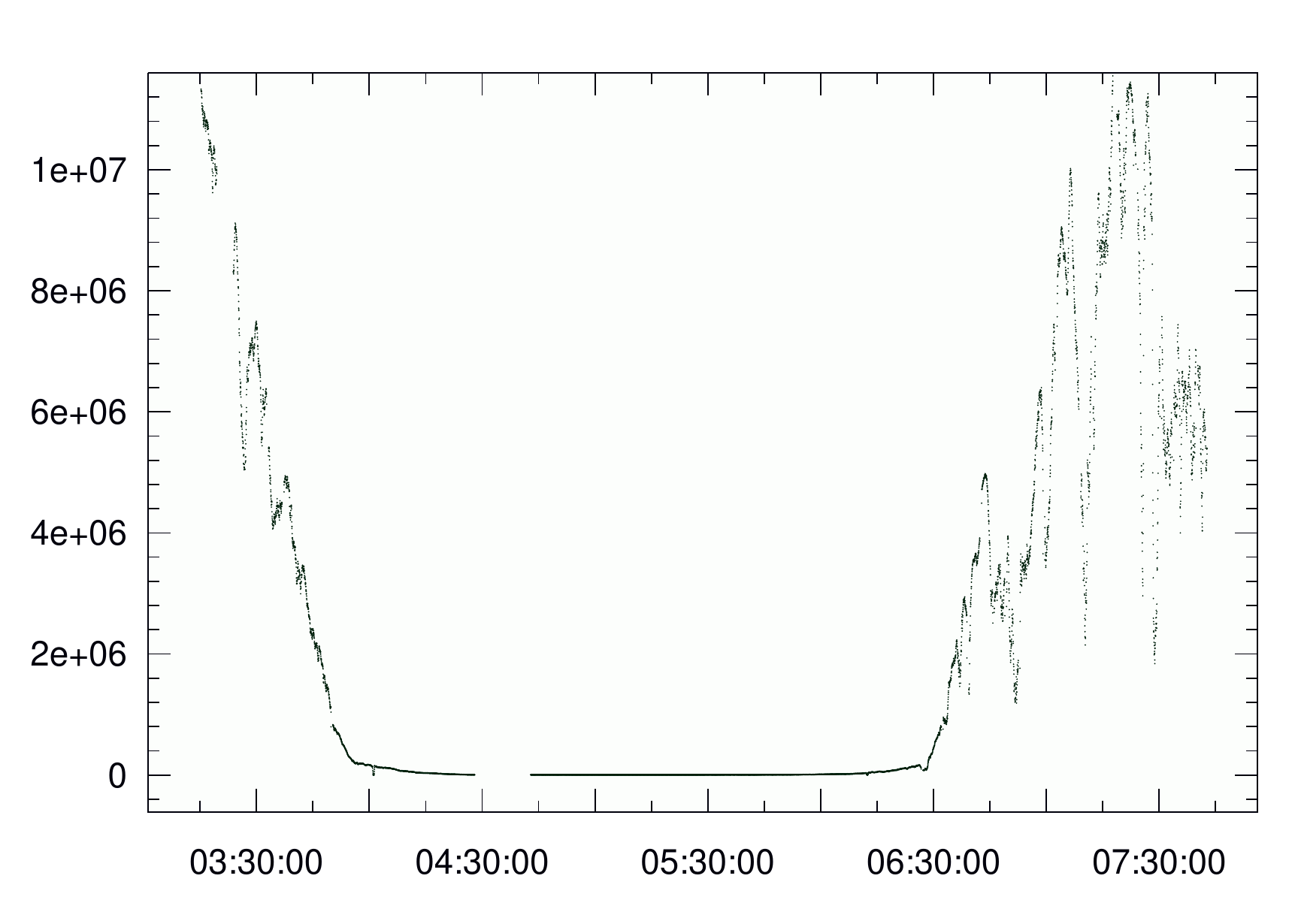}
\caption{\emph{a.} Initial fiber position on the Moon (central black dot). The 1.5\arcsec\ fiber is centered on the central mountain of Tycho. The size of the dot is to scale. Telescope tracking errors are such that the fiber drifted out of the crater by the end of the 4-hour time series. \emph{b.} Light curve from the fiber exposure meter. Time on the $x$-axis is UT. The $y$-axis is counts per second. The various dips are due to clouds passing through. }
 \label{F1}
\end{figure*}

Another approach to observe the spectral signature of Earth is during lunar eclipses (Pall\'e et al.~\cite{palle}, Vidal-Madjar et al.~\cite{vidal}, Garc\'ia Munoz et al.~\cite{garcia}, Ugolnikov et al.~\cite{ugo}, Arnold et al.~\cite{luc2}, Yan et al.~\cite{yan}, Kawauchi et al.~\cite{kawa}). The major difference to Earthshine observations is that in this case the radiation is dominated by selective absorption and scattering processes of sunlight transmitting the Earth atmosphere rather than by its reflection off the Earth's atmosphere, its clouds and the ground. The geometry of a lunar eclipse also resembles more that of an exoplanet transit in front of its host star than the classic Earthshine observation, and thus is of particular interest for the exoplanet community. In this way, Yan et al. (\cite{yan}) had detected the Rossiter-McLaughlin effect during the total lunar eclipse in April 2014. We note that the Earth transiting the Sun as seen from Jupiter's moons produced an inverse Rossiter-McLaughlin effect due to the opposition surge of icy Europa (Molaro et al. \cite{mol:bar}).  Arnold et al. (\cite{luc2}) had measured the Earth's atmosphere thickness as a function of wavelength while Garc\'ia Munoz et al.~(\cite{garcia}) had shown that the effective optical radius of the Earth during a transit is modulated by refraction (see also B\'etr\'emieux \& Kaltenegger \cite{bet:kal}).

Coyne \& Pellicori (\cite{coyne}) were the first to report the detection of a 2\%\ polarization degree of the integrated lunar disk during the eclipse on April\,13, 1968. Multiple scattering during the transmission through Earth's atmosphere was suggested as a possible cause of the polarization. After this first detection of polarization during lunar eclipses it took nearly 50 years until Coyne \& Pellicori's results were confirmed with modern observations. Takahashi et al. (\cite{tak}) used the 8.2\,m Subaru telescope to monitor the total lunar eclipse from April\,4, 2015. They observed polarization degrees of 2-3\,\%\ at wavelengths of 500$-$600\,nm and, in addition, an enhanced feature at the O$_2$ A-band near 760\,nm consistent with the explanation of polarization caused by double scattering during the first transmission through Earth's atmosphere. More recently, Takahashi et al. (\cite{tak2}) presented imaging polarimetry of the October 2014 lunar eclipse and found polarization degrees of less than 1\%. From a comparison of Earth's cloud coverage at the time of the two eclipses in 2014 and 2015, the authors found that the high cloud distribution was more inhomogeneous for the 2015 eclipse than it was for the 2014 eclipse. Recent radiative-transfer modeling results of the Earth's reflectance spectrum (Emde et al. \cite{emde}) showed that scattering in high ice water clouds and reflection from the ocean surface are crucial to explain the continuum polarization at longer wavelengths.

All previous spectro-polarimetric observations were made with low spectral resolution, $R=\lambda/\Delta\lambda$, of between 100--1000 in order to capture enough flux. However, Vidal-Madjar et al.~(\cite{vidal}) and Gonz\'alez-Merino et al. (\cite{sarg}) showed that spectral resolution is crucial for the detection of faint transmission species like sodium. The polarized radiative-transfer simulations of Earth's atmosphere by Emde et al. (\cite{emde}) and Stam (\cite{stam}) even suggest that there is still an overwhelming amount of detail in the polarized spectra waiting to be confirmed by observations. As far as we know, no high resolution polarimetric spectra of the Earthshine or inside or outside of a lunar eclipse were acquired until today.

In this paper, we present high resolution intensity as well as circular polarized (CP) and linear polarized (LP) spectra of the Moon during its eclipse January\,21, 2019. Our spectrograph is fiber-fed from two dual-beam polarimeters based on a Foster prism and a $\lambda/4$ retarder and thus laid out for observing point sources at a fixed spectral resolution of $R$=130\,000. Our spectra are always continuum rectified and the polarization defined by the difference of the ordinary and extra-ordinary beams and its sum. This in turn means that we are not recording continuum polarization by default, like all previous studies, but line polarization. The most important mechanism for CP and LP line polarization is the Zeeman effect while for LP line polarization coherent scattering effects like the Hanle effect provide a continuum for polarizing and de-polarizing spectral lines. It is particularly true for sunlight that CP and LP are much more prominent in individual (atomic) spectral lines than in the continuous spectrum (Trujillo Bueno et al. \cite{tru}, Stenflo \cite{sten}), opposite to reflective solar-system planets or any other diffuse sources like, for example, the interstellar medium. Resolving the individual lines of the oxygen bands and measuring their polarization levels may thus become a powerful technique to characterize exoplanet atmospheres with strong biomarkers. Importantly, stellar lines reflected from a planet become Doppler shifted according to the planet's (or in our case the Moon's) motion and are detectable with high spectral resolution (Martins et al.~\cite{mart}). Such periodic Doppler-shifts are thus expected from orbiting exoplanets and further help to disentangle telluric and exoplanetary signals.

Section~\ref{S2} in this paper describes our instrumentation and its setup for lunar observations. In Sect.~\ref{S3} we extract transmission  spectra from our time series that we then interpret as being a close proxy of a transmission spectrum of an Earth exoplanet. We also verify the Rossiter-McLaughlin effect for the January 2019 eclipse. Section~\ref{S4} is our paper summary. The Appendix gives the observing log and the radial-velocity (RV) data and some more technical details of the spectrum subtraction technique.

\begin{table}[!tbh]
\caption{Observing log for CD-III and CD-VI Stokes-I spectra on UT January 21, 2019. Eclipse phases are for crater Tycho.}\label{T1}
\begin{flushleft}
\begin{tabular}{llll}
\hline\hline\noalign{\smallskip}
 UT start   & $t_{\rm exp}$ & \multicolumn{2}{c}{$<S/N>$}  \\
 (hh:mm:ss) & (mm:ss)       &  CD\,III & CD\,VI \\
\noalign{\smallskip}\hline\noalign{\smallskip}
penumbral ingress:& & & \\
 03:24:57.4 & 00:15.000 &  712  &  907  \\
 03:27:06.6 & 00:20.004 &  727  &  834  \\
 03:28:56.4 & 00:20.000 &  788  &  937 \\
 03:30:57.5 & 00:20.000 &  712  &  866 \\
 03:33:27.7 & 00:20.000 &  600  &  768 \\
 03:35:10.1 & 00:20.000 &  564  &  711 \\
 umbral phase:& & & \\
 04:04:01.2 & 06:00.000 &  186  &  355 \\
 04:11:22.5 & 06:00.000 &   91  &  216 \\
 04:46:01.4 & 20:00.000 &    1  &  102 \\
 05:07:23.6 & 20:00.000 &    1  &   84 \\
 05:29:16.8 & 20:00.000 &    1  &  182 \\
 05:50:30.5 & 20:00.000 &   62  &  329 \\
 06:12:51.6 & 02:00.000 &   39  &  119 \\
 06:16:12.9 & 02:00.000 &   50  &  139 \\
 06:19:31.0 & 01:56.294 &   78  &  170 \\
 06:22:48.9 & 02:38.106 &  141  &  244 \\
 06:27:19.0 & 01:51.436 &  140  &  245 \\
 penumbral egress:& & & \\
 06:30:23.8 & 00:30.417 &  142  &  255 \\
 06:53:14.2 & 00:10.125 &  236  &  343 \\
 06:54:45.7 & 00:08.612 &  241  &  343 \\
 06:56:12.6 & 00:07.686 &  262  &  370 \\
 06:57:43.4 & 00:05.602 &  278  &  401 \\
 06:59:42.0 & 00:08.682 &  243  &  364 \\
 07:01:05.0 & 00:05.598 &  263  &  399 \\
 out-of-eclipse: & & & \\
 07:24:28.1 & 00:03.078 &  280  &  400 \\
 07:25:52.8 & 00:02.554 &  250  &  363 \\
 07:27:13.3 & 00:03.097 &  310  &  450 \\
 07:28:38.0 & 00:16.781 &  251  &  331 \\
 07:30:48.0 & 00:05.135 &  298  &  450 \\
 07:32:07.6 & 00:05.146 &  244  &  387 \\
\noalign{\smallskip}
\hline
\end{tabular}
\tablefoot{The time series in CD-VI was interrupted several times to obtain spectra in other CD combinations. On UT 3:37-3:53, 6:32-6:40 and 7:03-7:10 for CD II+IV; and on UT 6:43-6:50 and 7:12-7:20 for CD III+V. }
\end{flushleft}
\end{table}

\section{Observations}\label{S2}

\subsection{Instrumental set up}

All observations were obtained with PEPSI (Strassmeier et al.~\cite{pepsi}) at the 2$\times$8.4\,m LBT (Hill et al.~\cite{lbt}) in Arizona. Each of the two 8.4-m telescopes carries one polarimeter in their symmetric straight-through Gregorian foci. We used the two 8.4\,m LBT mirrors dubbed SX and DX always in binocular mode. Two pairs of octagonal 200$\mu$m core-diameter fibers per polarimeter feed the ordinary and extraordinary polarized beams per telescope via a five-slice image slicer per fiber into the spectrograph. The fiber aperture projected on the sky has a circular diameter of 1.5\arcsec . The two pairs of fiber provide then four spectra per \'echelle order on the CCD. Each spectrum has a resolution of $R\approx$130\,000 (0.06\,\AA\ or 6\,pm at 7600\,\AA) and is being sampled by 4.2 CCD pixels. PEPSI has a blue and a red-optimized arm, each with three cross dispersers (CD), two of them simultaneously. We focussed our time series on the O$_2$ A-band within CD\,VI but kept CD\,III in the blue arm.

Both polarimeters are based on a classical dual-beam design with a modified Foster prism as linear polarizer with two orthogonally polarized beams (ordinary and extra-ordinary) exiting in parallel. The unit holding the polarimetric optics, that contains the Foster prism, the atmospheric dispersion corrector (ADC) behind the Foster, the two fiber heads, and the two fiber viewing cameras is rotatable with respect to the parallactic axis on the sky. A quarter-wave retarder for circular polarization is inserted into the optical beam in front of the Foster prism for the CP measurements and retracted for the LP measurements. There is no half-wave retarder. This design minimizes the cross-talk between CP and LP to less than 0.1\%\ (Ilyin \cite{ilya12}). However, the polarimeters cannot record sky polarization simultaneously because PEPSI's sky fibers are used for the additional orthogonal beams. Previous low resolution observations used a long-slit or a multi-slit set-up in which one end of the slit, or one of the slits, was positioned just outside the Moon and thus record sky, which then can be used for telluric subtraction. Several authors proceeded in this way for intensity and polarimetric observations (e.g., Pall\'e et al.~\cite{palle}, Sterzik et al. \cite{sterz}, Takahashi et al. \cite{tak}). High-resolution fiber-fed spectrographs usually can not do this because their fibers are too close together on the sky, in particular PEPSI cannot do this if used in spectro-polarimetric mode. The spectrograph and the polarimeters were described in more detail in Strassmeier et al. (\cite{pepsi}, \cite{spie-austin}).

\begin{figure*}
\includegraphics[angle=0,width=\textwidth,clip]{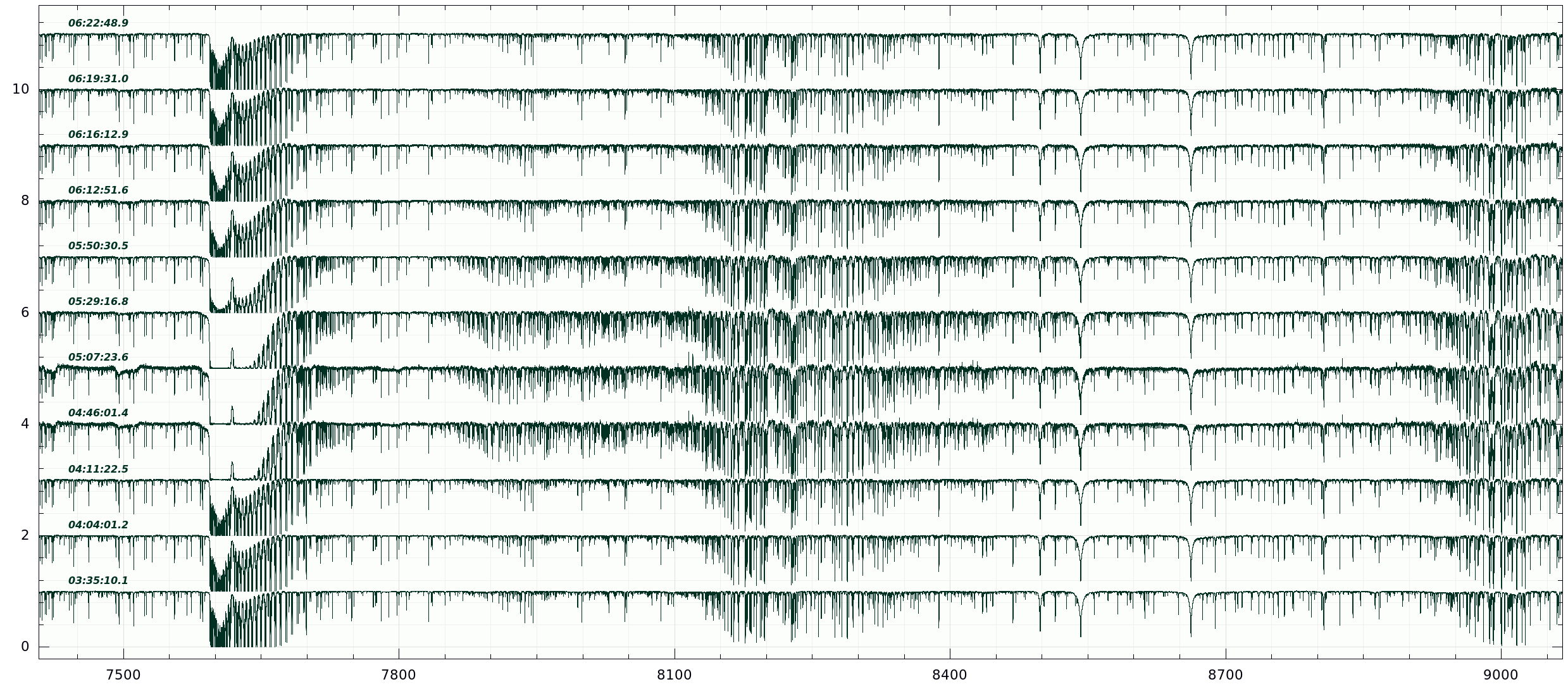}
\caption{Time series of Stokes-I spectra around umbral eclipse. Wavelengths are given in \AA ngstroem, individual spectra are offset by 1.0 in intensity for better viewing. Shown is the entire wavelength range of CD\,VI. Notice the deepening of the O$_2$ A-band absorption at 7600\,\AA\ during umbral eclipse between UT\,4:46 and 5:50. Its absorption became so dominant that no light at these wavelengths was left except a residual emission feature at 7618\,\AA . Increased absorption line strengths of other molecular bands are also seen, most notably for the H$_2$O bands between 7850--8450\,\AA . }
 \label{F2}
\end{figure*}

\subsection{Data acquisition}

Observations of the Moon commenced between UT\,3:25 and 7:41 on January 21, 2019. Maximum of the total eclipse was at UT\,5:12. Sky conditions were moderate but sky was clear at the beginning of the night with increasing cirrus until the end of the observations at around midnight. We estimated a 2-mag extinction at the end of the eclipse due to increased cirrus. Because we observe the ordinary and extra-ordinary beams simultaneously the cirrus affects the polarimetric accuracy mainly through the increased sky polarization and its possible variation within an exposure but also due to the related lowered signal-to-noise ratio (S/N). The visual binary HIP\,116035 was used for north-south alignment of the Foster prism. After centering on a nearby guide star for an encoder reset the center of the Tycho crater of the Moon was acquired using its JPL Horizons online ephemeris\footnote{Jet Propulsion Laboratory: http://ssd.jpl.nasa.gov/}. The selenographic coordinates were E-Long 348.80 and Lat --43.30 degrees. At that time the Moon was at a height above the horizon, $h$, of just 25\degr . Penumbral eclipse started already before Moon rise while the (global) umbral eclipse started at UT\,3:34. Total umbral eclipse began at UT\,4:41 and lasted until 5:43. Penumbral egress lasted until UT\,7:22.

\begin{figure}
\includegraphics[angle=0,width=86mm,clip]{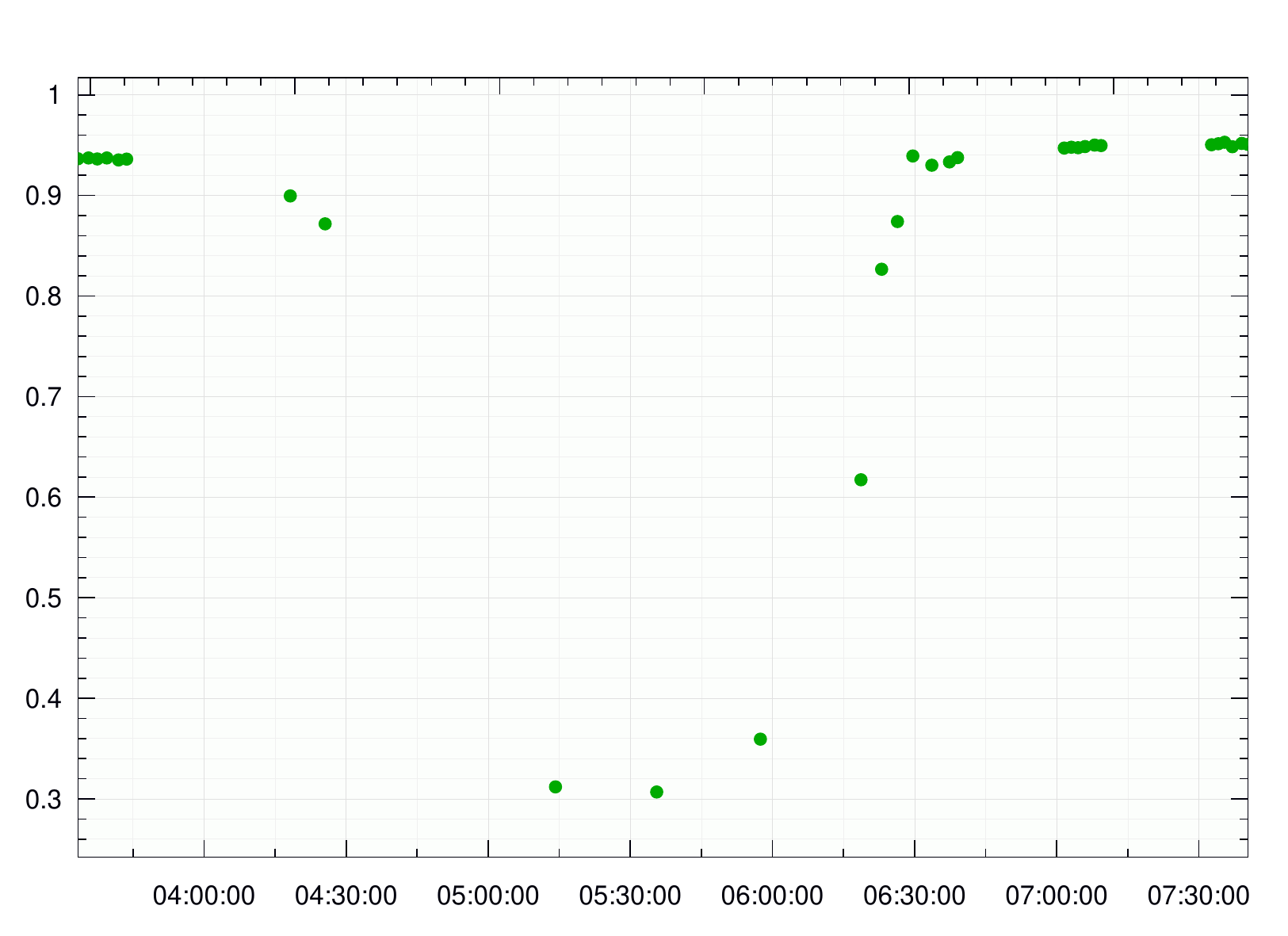}
\caption{Strength of the residual O$_2$ A-band emission feature at 7618\,\AA\ throughout the time series. A residual intensity of unity is the continuum level. Time on the $x$-axis is UT for January 21, 2019. }
 \label{F3}
\end{figure}

The Tycho region entered Earth's shadow at UT\,4:00 when the Moon was at $h$=40\degr\ and reappeared in sunlight at UT\,6:26 when at $h$=72\degr . We note again that the fiber aperture projected on the Moon is just 1.5\arcsec . Thus, the spectrograph sees light from only a geometrically tiny fraction of the Moon's surface, but see Qiu et al. (\cite{qiu}) for contamination from stray light from one point of the Moon to others. In fact, the apparent diameter of the Moon during the eclipse was $\approx$0.56\degr\ and the apparent diameter of the Tycho crater is thus $\approx$50\arcsec\ (86\,km on the Moon). Our fiber was positioned on the central mountain of Tycho which appears as a bright spot in white-light images of the Moon. Figure~\ref{F1}a shows the Tycho crater and the position and diameter of our fiber aperture to scale. We note that the PEPSI polarimeters each have an entrance diaphragm wheel with several pinholes that can be chosen according to the type of target. Our default diaphragm was the smallest available in order to minimize the contribution from sky polarization and is 5\arcsec\ in diameter. The two fibers per polarimeter have their own fiber aperture diaphragms. These match the fiber core diameter and set the fiber field of view to 1.5\arcsec\ (of the 5\arcsec\ available). LBT was run in binocular mode, which means both telescopes are collimated through their respective wavefront sensors onto the same target. Thus, our spectrograph sees light from four fibers pointing at the same location on the Moon. Telescope guiding was disabled and replaced by non-sidereal tracking. Tracking errors are such that we can expect a maximum drift rate of 0.3\arcsec\ per minute. This converts to a maximum scan length on the Moon of 6\arcsec\ for the long exposures during umbral eclipse and negligible for the others. In total, we estimate that our fiber had drifted out of the crater Tycho by the end of the 4-hour time series.

PEPSI runs exposure meters for each of the two telescopes. Two gray beam splitters in front of the image slicers divert less than 1\%\ of the light into two Sens-Tech P25232 R1925A photomultiplier tubes of S20-type response (optimized for wavelengths between 3000--6000\,\AA ). They record the photon flux collected by the telescopes and fiber-core diameters projected on the sky. In our case, we were recording the brightness of the Moon in the 1.5\arcsec\ fiber aperture centered on the crater Tycho. Figure~\ref{F1}b shows the light curve of the Moon from the DX channel. The count rates are not calibrated as for classical photometry but the drop from 1.2$\times$10$^7$~cts/s during outside of eclipse to $\approx$600~cts/s during mid totality corresponds to a drop of 10\fm75 in approximately white light.

Stokes I was built from the sum of the ordinary plus the extra-ordinary beams as well as from SX and DX. Thirty Stokes-I spectra were available in two wavelength regions with cross disperser III covering 4800--5441~\AA\ and CD~VI covering 7419--9067~\AA\ (Table~\ref{T1}). Individual exposures were set for retarder angles of 45\degr\ and 135\degr\ for Stokes V, Foster prism position angles of 0\degr\ and 90\degr\ for Stokes Q and 45\degr\ and 135\degr\ for Stokes U with respect to the north-south direction. One polarimetric exposure cycle consisted then of six sub-integrations. We note again that the $\lambda/4$ retarder was removed from the beam during the LP exposures. Exposure time per sub-integration was between 3--16\,s outside of eclipse, between 20--360\,s during penumbral eclipse, and 20\,min during umbral eclipse. Median S/N per pixel over the entire cross-disperser wavelength range outside of eclipse was around 750 in CD~VI (peak of 1000) and 550 in CD~III (peak of 750). However, S/N was only 85 and 3 during umbral eclipse for the two wavelength ranges respectively, that is basically no blue light was received during umbral eclipse. Only one independent exposure cycle of four subintegrations was possible during umbral eclipse\footnote{A fifth exposure was obtained at UT\,4:33 at the very beginning of the umbral eclipse but was lost due to a multi-threading problem in the X11 software (the problem later was fixed).} of which we derived four Stokes-I spectra, one Stokes-Q, one Stokes-U, and one Stokes-V spectrum, always normalized to the continuum intensity $I_{\rm c}$. A few more spectra were obtained with the cross disperser combinations II+IV and III+V. The log of all individual sub-exposures for CD\,VI is given in the Appendix in Table~\ref{T1-App}. The radial velocities are given in Table~\ref{T2-App}.

\begin{figure*}
\center
\includegraphics[angle=0,width=120mm,clip]{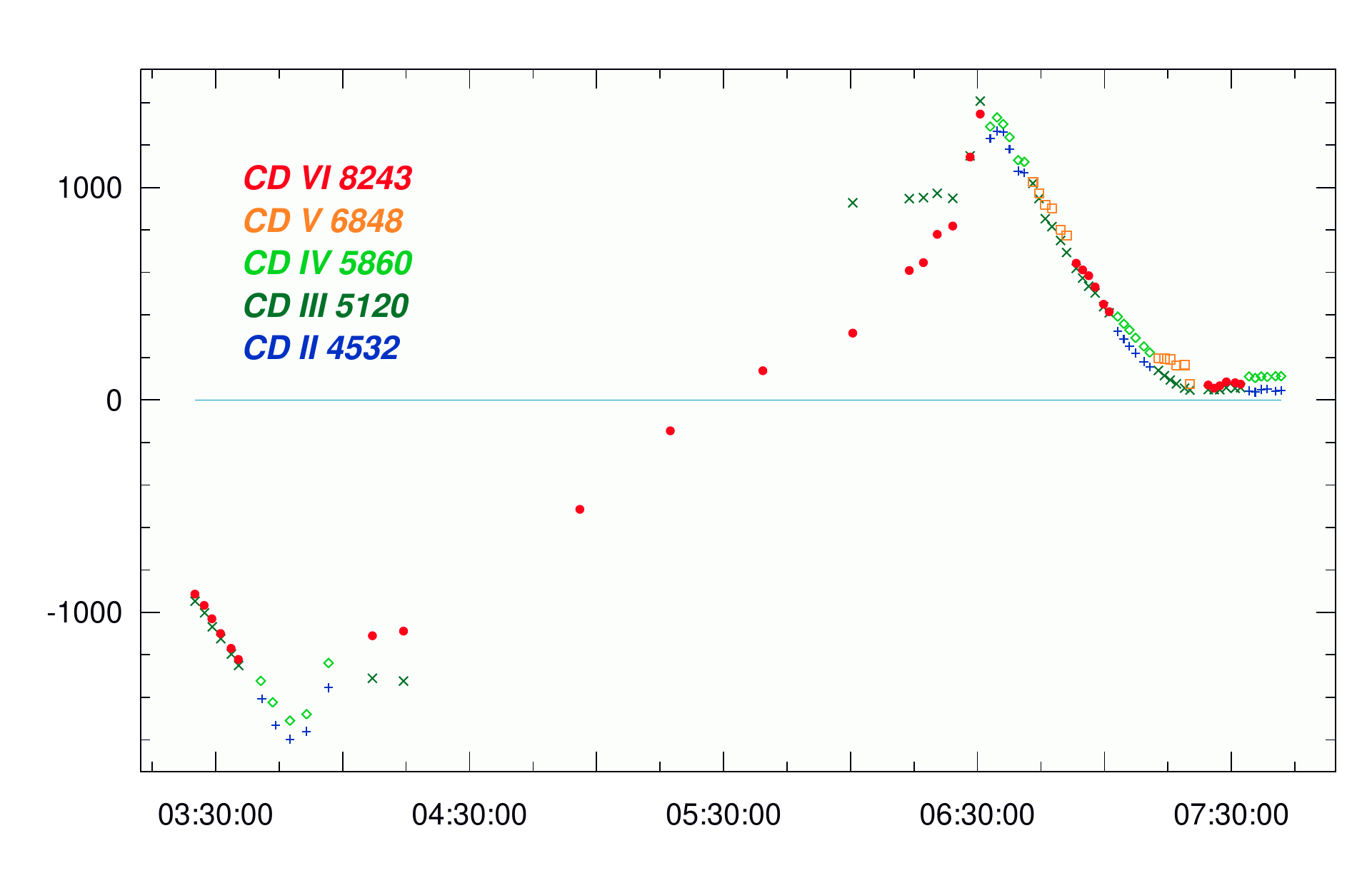}
\caption{Heliocentric radial velocities in \ms\ from the lunar Tycho region. The dots are our relative velocities from solar absorption lines with respect to the IAG FTS solar atlas (Reiners et al. \cite{iagfts}). The colors denote the wavelengths at which the velocities were measured and are identified in the insert (CD is the cross disperser and its respective central wavelength in \AA ; symbols are dots = CD\,VI, squares = CD\,V, diamonds = CD\,IV, crosses = CD\,III, pluses = CD\,II). The observed velocities trace the Rossiter-McLaughlin effect from Earth transiting in front of the Sun as seen from Tycho on the Moon.}
 \label{F4}
\end{figure*}

\subsection{Data reduction and calibration}

Data reduction was done with the software package SDS4PEPSI (``Spectroscopic Data Systems for PEPSI'') based on Ilyin (\cite{4A}), and described in some detail for integral-light spectra in Strassmeier et al. (\cite{sun}). The specific steps of image processing are the same for polarimetric spectra and include bias subtraction and variance estimation of the source images, super-master flat field correction for the CCD spatial noise, scattered light detection from the inter-order \'echelle regions and subtraction, definition of \'echelle orders, wavelength solution for the Th-Ar images, optimal extraction of image slicers and cosmic spikes elimination, normalization to the master flat field spectrum to remove CCD fringes and the blaze function, a global 2D fit to the continuum, and the rectification of all spectral orders into a 1D spectrum. The scattered light detection is based on the inter-order CCD space and then subtracted from the CCD pixels that cover the orders. Typical scattered-light values were 0.5\%\ in CD\,VI, 1\%\ in CD V--III, and 1.5\%\ in CD\,II.

The wavelength calibration is done for each image slicer separately by forming a 3D Chebyshev polynomial fit of each Th-Ar line position to its normalized wavelength, order number, and slice number. A typical error of the fit in the central part of the image is 3--5\,\ms . The basic polarimetric calibration procedure was done in two main steps. Firstly, aligning of the optical axis of the Glan-Thompson polarizing prism in the calibration unit with respect to the Foster beam-splitting prism and, secondly, aligning of the optical axis of the science retarder with respect to the polarizer. Details were given by Ilyin et al. (\cite{florence}).

\section{Analysis}\label{S3}

\subsection{Tracing the eclipse}

Figure~\ref{F2} shows a time series of 11 Stokes-I spectra in CD\,VI centered around the umbral eclipse. The increase of molecular O$_2$ absorption during the eclipse is easily noticeable by eye. The O$_2$ A-band got so strong and broad that basically no solar flux was transmitted and the intensity went to practically zero. Only an intermittent pseudo emission line at 7618\,\AA\ remained. We measured the residual intensity of its peak, dubbed an emission-line strength, and plot the value versus time in Fig.~\ref{F3}. The feature is not an emission line in the physical sense but is artificially produced by increased line absorption in the blue (R-branch) as well as in the red (P-branch) part of the A-band. Because the actual O$_2$ lines are mostly saturated and thus do not carry a reliable flux signal any more, this feature can be used to trace the eclipse. In geophysics, the A-band characteristics are among the main tracers for aerosol and cloud optical properties (e.g., Pitts et al.~\cite{pitts}, Emde et al. \cite{emde}).

In the following, we first identify and then isolate the various spectrum components in our high-resolution spectra and then search for other chemical species in the Earth's transmission spectrum.

\subsection{Identifying the spectrum components}

Each spectrum in Fig.~\ref{F2} is the combination of three contributing components. The first component is the Sun and its black-body continuum with an atomic absorption line spectrum as we know it from direct solar observations (e.g., Kurucz et al.~\cite{nso}) or from our own PEPSI Sun-as-a-star atlas (Strassmeier et al.~\cite{sun}). The effective temperature of the solar photosphere also allows molecular line formation, in particular above the atmosphere of cool sunspots. The most prominent molecular species at optical wavelengths are TiO, VO, and CO. We assume the solar source remained constant during the four hours of our eclipse observations, that is its intrinsic line depths remained constant. Solar spectral lines will follow the radial velocity as seen from Earth and will change in the four hours of the eclipse observations. We measure these velocities from wavelength regions where no telluric lines are present. Such a telluric-free zone is just blueward of the O$_2$ A-band between 7517--7580\,\AA\ and contains over 20 well-defined and relatively unblended solar lines. We adopt the IAG FTS atlas from Reiners et al. (\cite{iagfts}) as a template and cross correlate all other spectra with it. The cross-correlation peak is fit with a polynomial and returns one relative velocity. Heliocentric and diurnal corrections for the times of our data are obtained with the JPL Horizons calculator (Sect.~2.2) and applied as well. The resulting heliocentric radial velocities are plotted in Fig.~\ref{F4} and are listed in the Appendix in Table~\ref{T2-App} along with other relevant information. These velocities trace the eclipsed Sun (by the Earth) as seen from the crater Tycho on the Moon. The Earth blocked that part of the solar disk that rotated toward  the observer in Tycho at ingress and then that part that rotated away from the observer in Tycho at egress. Its approximate sinusoidal shape is the analog of a Rossiter-McLaughlin effect of a transiting exoplanet (e.g., Triaud \cite{triaud}, Yan et al. \cite{yan}, Dreizler et al.~\cite{drei}). Because we look at different heights and convective velocities in the solar atmosphere at the solar limb when using different wavelength regions for the cross correlation, the resulting RVs from blue wavelengths are larger by $\approx$50\,\ms\ to the ones from red wavelengths at the times the solar light is predominantly from the limb. This is particularly noticeable in Fig.~\ref{F4} for the blue CD~III RVs at around UT~04:15 and UT~06:15 with respect to the red CD-VI velocities. We also note that a RV offset of 74\,\ms\ remained after the end of the eclipse and is just due to an unaccounted zero point in the IAG FTS atlas with respect to PEPSI.

\begin{figure*}
{\bf a. Ingress at UT\,3:50} \hspace{63mm} {\bf b. Egress at UT\,6:30}\\
\includegraphics[angle=180,width=86mm,clip]{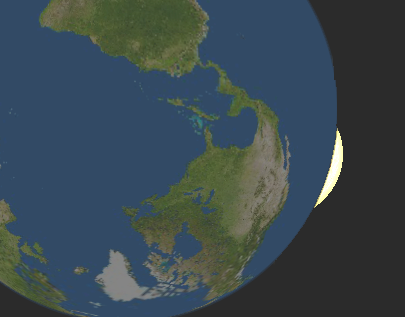}\hspace{5mm}
\includegraphics[angle=180,width=88.5mm,clip]{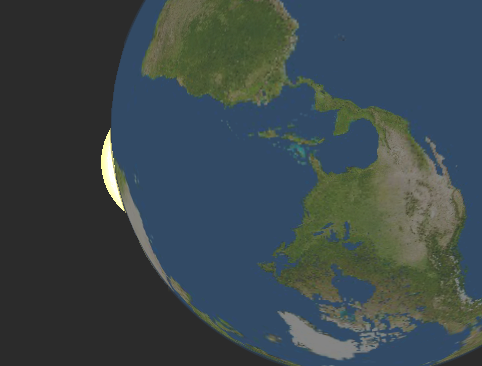}
\caption{The January~21, 2019 eclipse as seen from the Moon's crater Tycho. \emph{a.} During deep penumbral phase UT\,3:50 at ingress; \emph{b.} during deep penumbral phase UT\,6:30 at egress. The ingress phase happened when the Earth's limb was seen with its water surface in front of the Sun, the egress when mostly Africa was in front of the Sun. Plot program is Redshift-6 (2014 United Soft Media Verlag GmbH). }
 \label{Fmaps}
\end{figure*}

The second spectrum component is the Earth itself. Its spectrum is more complex and is built from two transmissions of the solar light through different regions in our atmosphere and from opposite directions. The first transmission is the direct sunlight passing through the atmosphere toward the Moon. This is the component we are interested in. The Moon ``sees'' the Earth in front of the Sun with our atmosphere sunlit so that it appears like the solar corona during a total solar eclipse except that the terrestrial disk appears much larger than the solar disk. Figure~\ref{Fmaps} illustrates how the eclipse geometry appeared as seen from Tycho. The Sun was setting with the northern Pacific in front at coordinates 32$^\circ$42\arcmin N, 145$^\circ$10\arcmin W, and was rising with northern Africa in front at coordinates 28$^\circ$28\arcmin N, 03$^\circ$12\arcmin W. The cloud fraction observed by Aqua/MODIS\footnote{http://modis.gsfc.nasa.gov/ and https://neo.sci.gsfc.nasa.gov/} for these locations on Jan.~21, 2019 was greater than $\approx$60\% in the northern Pacific and less than 5\% in northern Africa. The dominant radiative-transfer mechanisms are selective molecular absorption, Rayleigh scattering as well as refraction. The $\lambda^{-4}$ dependence of Rayleigh scattering effectively keeps the blue light from reaching the Moon during eclipse. It is the reason why the S/N of our blue CD\,III spectra basically went to zero. The remaining red light undergoes refraction by the Earth's atmosphere and focusses onto the Moon like in a spyglass (that's why the eclipsed Moon appears bloody red). The selective absorption must be rather inhomogeneously distributed around the atmospheric circumference because our atmosphere is structured differently above sea and land, and poles and equator as well as with height above ground (see Kawauchi et al.~\cite{kawa}). The rotation of the Earth causes some of these absorption lines appear red shifted, some blue shifted, and some not shifted at all, depending where in the Earth's atmosphere the sunlight had been absorbed.  Therefore, a residual spectrum over a large wavelength range may appear with conflicting radial velocities and it is important to isolate the wavelength regions of interest. The maximum expected shift is the equatorial rotational velocity of Earth of $\pm$460\,\ms .

The second path through the Earth atmosphere, called the second transmission, takes place when the Moon reflects the first transmission and then enters the Earth atmosphere above our telescope. It is selectively absorbed along its path by the constituents of Earth's nighttime atmosphere, most notably again by molecular oxygen and water vapor. These telluric lines  basically follow the Earth's rotational velocity above our telescope and thus do not change in radial velocity but change in intensity due to the variation in air mass. Some residual RV changes are nevertheless introduced by local velocity fields in the Earth's atmosphere above the telescope.

The third component is the Moon. For our purpose it is just a reflector. However, our fiber is positioned at the Tycho location on the Moon on its southern hemisphere and not directly at its disk center. Fortunately, the bound rotation of the Moon does not introduce much of an extra velocity component. We compute the expected additional radial velocity due to the Moon's orbital motion at the position of Tycho with respect to our geographic coordinates using the JPL Horizons calculator and get $\approx$3~\ms . Albeit very small compared to the diurnal or heliocentric corrections it is included in the corrections in Fig.~\ref{F4}. Besides this extra radial-velocity shift, we see no other traces of the Moon in our spectra. We note that the bright (higher albedo) regions on the Moon depolarize the continuum light more strongly than the dark (lower albedo) regions. This is known since Dollfus (\cite{doll}) and is related to the different porosity and the average size of particles of the lunar soil in mare as compared to highlands and craters (see later in Sect.~\ref{S34}). Our instrument's field of view always remained in regions of high albedo.

\subsection{Isolating the spectrum components}\label{S33}

The next step is to separate the three components and isolate the transmission spectrum. We apply two commonly used methods from exoplanet transmission spectroscopy or binary-spectrum decomposition. One method is based on the subtraction of a reference out-of-eclipse spectrum, the other on its division.

\begin{figure*}
\includegraphics[angle=0,width=\textwidth,clip]{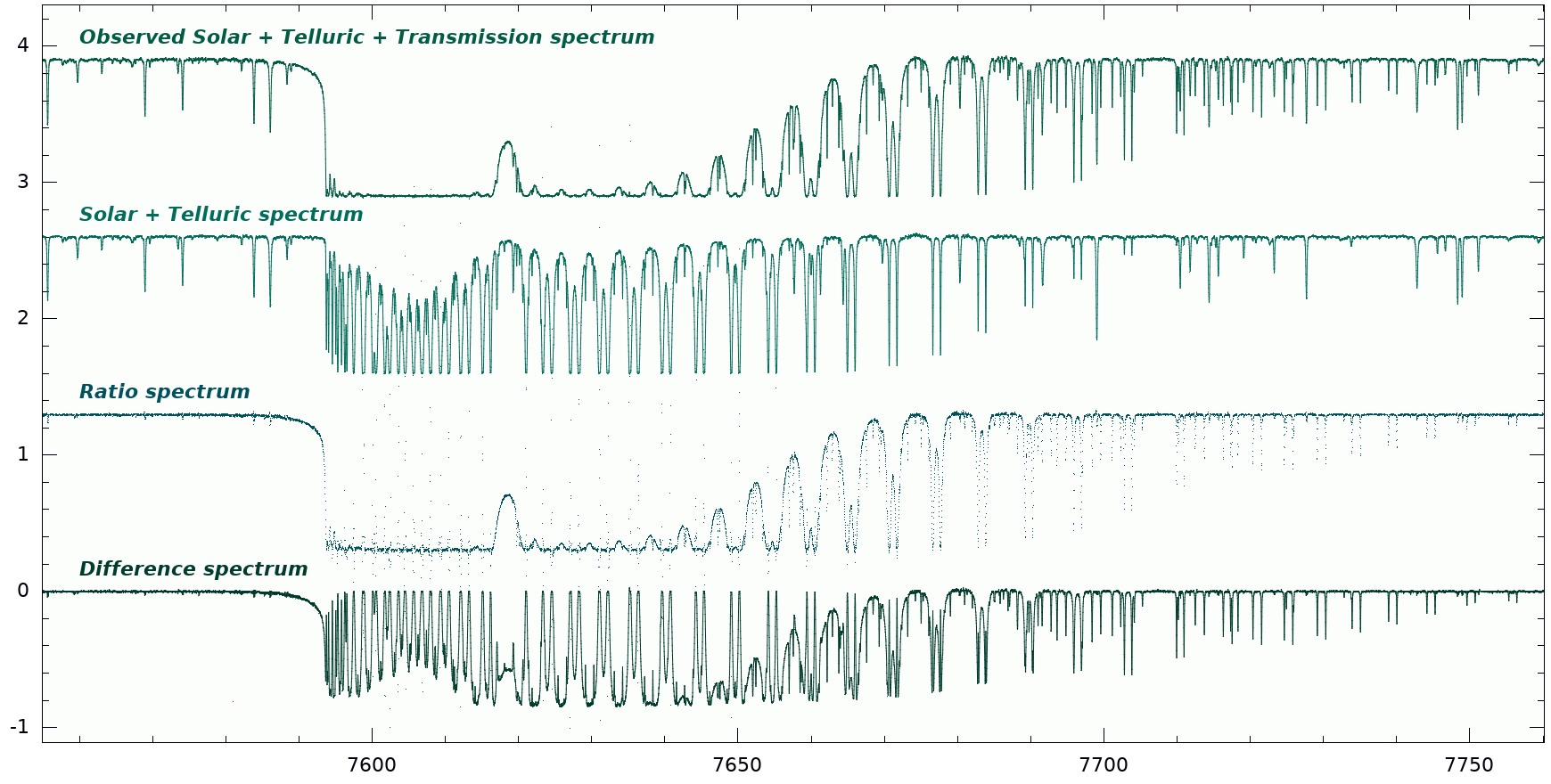}
\caption{A comparison of the spectrum components in the O$_2$ A-band. The top spectrum is the observed spectrum during umbral eclipse with all components in it. The next lower spectrum is the solar plus the telluric component obtained from a smoothing-spline fit to all spectra versus air mass and time. The two bottom spectra are the ratio and the difference of above, respectively, and reveal the Earth's transmission spectrum. A vertical offset of 1.3 is applied for better visibility. Wavelengths are in \AA ngstroem. We note that the ratio spectrum is plotted with dots in order not to overemphasize the many divisions by nearly zero.}
 \label{F6}
\end{figure*}

\emph{Method 1}. Difference spectra are build with respect to averaged post-transit exposures (see, e.g., Aronson \& Wald\'en \cite{aro:wal}). Because of our high spectral resolution and S/N, we expect seeing line-depth and RV changes of the telluric lines caused by changes of the local atmosphere over the coarse of the eclipse. Therefore, we first create interpolated out-of-eclipse spectra and then subtract them from each individual observation. In detail, we built residual spectra by rectifying the intensity deviation per pixel in wavelength as a function of time. A low order smoothing spline is used to fit each pixel intensity. It is like a spectral continuum fit in time: the residuals of the fit exclude all points which are not changing in time (like all solar lines or wavelength regions unaffected by telluric lines, e.g., near $\lambda$7550\,\AA ) but retain points which are changing in time due to eclipse. The fit is then subtracted from each observed spectrum in the time series. Figure~\ref{F1App} in the Appendix shows a typical fit's relative intensity as a function of time (for a particular pixel within a telluric line of appreciable depth as well as outside of a telluric line). This procedure not only effectively removes the solar spectrum because it is comparably constant but also removes the air-mass dependent telluric spectrum from the second transmission without removing the extra changes during the eclipse (a higher order spline would also remove the eclipse). Air mass changed from 1.7 at the beginning to 1.0 at the end of our observations. The remaining residual (difference) spectra are ideally zero for the phases outside eclipse if no further contributing sources were present or if the Earth's nighttime atmosphere were perfectly constant. Whatever was imprinted during the first transmission would remain in the eclipse spectra, at least in first order. Figure~\ref{F2App} in the Appendix is a plot of all residual (difference) spectra as a function of time. Figure~\ref{F6} is a zoom into the O$_2$ A-band and shows the averaged residual mid-umbral spectrum (averaged from the spectra at UT\,04:46, 05:07, and 05:29), and also the subtracted solar and telluric spectra for comparison. The averaged umbral spectrum is the total transmission spectrum of the Earth's atmosphere as seen from an observer sitting in the Tycho crater on the Moon and enjoying a total solar eclipse.

\emph{Method 2}. The more simple technique is building spectral ratios (see, e.g., Cauley et al. \cite{cau}). In our case we use the above mentioned spline fit also for division. In order to reach higher S/N during umbral phase, we use the average of the three observed CD-VI spectra between UT~4:46 and 5:29 and divide them by the average of the three out-of-eclipse spectra at around UT~7:30. Otherwise, individual spectra are used. Figure~\ref{F3App} shows the Na\,D wavelength region from the full log of CD-IV spectra as an example. The ratio method works fine as long as there is no division by zero or nearly zero. For the O$_2$ A-band it is not optimal to divide the individual spectra by an outside-eclipse spectrum because the A-band line flux during eclipse is practically zero (see Fig.~\ref{F2}). Even the out-of-eclipse spectra have most of its A-band lines almost completely saturated and thus at nearly zero intensity.

Figure~\ref{F6} compares all spectrum components for the A-band wavelength region. The two lower spectra are the respective difference and ratio spectra from the two methods. Because the atmospheric refraction component in the umbral-eclipse spectrum would be different in case of an exoplanet-transit observation (see Garc\'ia Munoz et al.~\cite{garcia} and B\'etr\'emieux \& Kaltenegger \cite{bet:kal}), we consider our transmission spectra still just approximations for the generalized stellar transit case. The main difference between our transmission spectrum and a standard telluric spectrum (with the Sun or bright stars as light source) is the geometric path of the photons through the Earth's atmosphere. A spectrum during a lunar eclipse is thus a better approximation because it comes from the same path as if we would observe a transit of an exoplanet, while the standard-star spectra are integrations along a pencil beam usually perpendicular to the surface (if at air mass 1.0) and either through the daytime atmosphere or the nighttime atmosphere.

\begin{figure}
\center
\includegraphics[angle=0,width=87mm,clip]{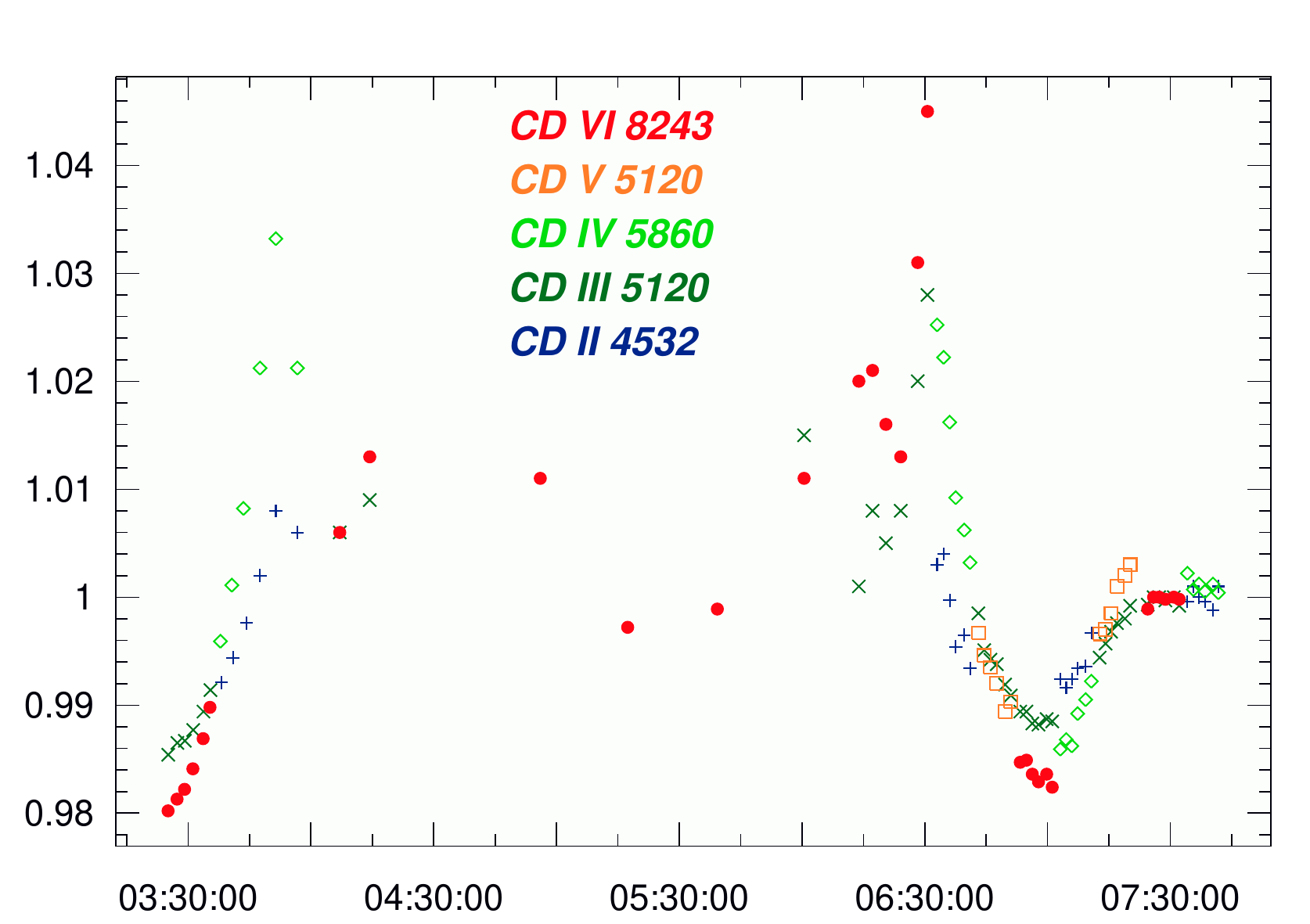}
\caption{Combined CLV and Ring correction as a function of time for all PEPSI wavelength ranges. The color and symbol code indicates the cross dispersers and their respective central wavelength in \AA\ and is the same as in Fig.~\ref{F4}. The $x$-axis is time on January~21 in UT, the $y$-axis is the dimensionless combined correction factor. The umbral eclipse for Tycho lasted from UT\,4:00 to UT\,6:26.}
 \label{Fring}
\end{figure}

\subsection{Searching for telluric atomic line absorbers}

\subsubsection{Removing the combined effect of scattering and center-to-limb variation}

From a single spectrum during the penumbral phase of the partial eclipse in August 2008, Vidal-Madjar et al. (\cite{vidal}) succeeded in the detection of telluric molecular oxygen and ozone, but also of neutral sodium. They employed the SOPHIE instrument with a spectral resolution of 75\,000 and were followed up with HARPS and UVES at $R\approx 115\,000$ for the eclipse in December 2010 (Arnold et al. \cite{luc2}). The only other atomic-line detection presented so far, that is for Ca\,{\sc ii} H\&K and the infrared triplet by Pall\'e et al.~(\cite{palle}), was at a resolution of only $\approx$1\,000 and was likely an artifact (see the discussion in Vidal-Madjar et al. \cite{vidal}). Arnold et al. (\cite{luc2}) re-emphasized the Ring effect and its impact for eclipse  observations, that is that the depth of solar Fraunhofer lines from scattered light is lower than from direct sunlight (Grainger \& Ring \cite{gra:rin}), mainly due to rotational Raman and forward scattering in the Earth's atmosphere (see also Garc\'ia Munoz et al.~\cite{garcia}). Vidal-Madjar et al. (\cite{vidal}) and later Arnold et al. (\cite{luc2}) tested various methods to mitigate this effect. The simplest is removing a small continuum constant from the observed spectra (first applied by Noxon et al. \cite{nox}) in order to mimic the expected lower line depths. The correction depends on the eclipse phase in the sense the deeper the Earth's atmosphere is penetrated the higher the constant, but also depends on wavelength with the redder lines being less affected by a factor $\approx$3 than the bluer lines (e.g., Pallamraju et al. \cite{palla}). Expected correction values are approximately between 4\%\ and 0.6\%\ for the deepest and the shallower phases of penumbral eclipse, respectively.

\begin{figure*}
{\bf a.} \hspace{90mm} {\bf b.} \hspace{40mm} {\bf c.}\\
\includegraphics[angle=0,width=92mm,clip]{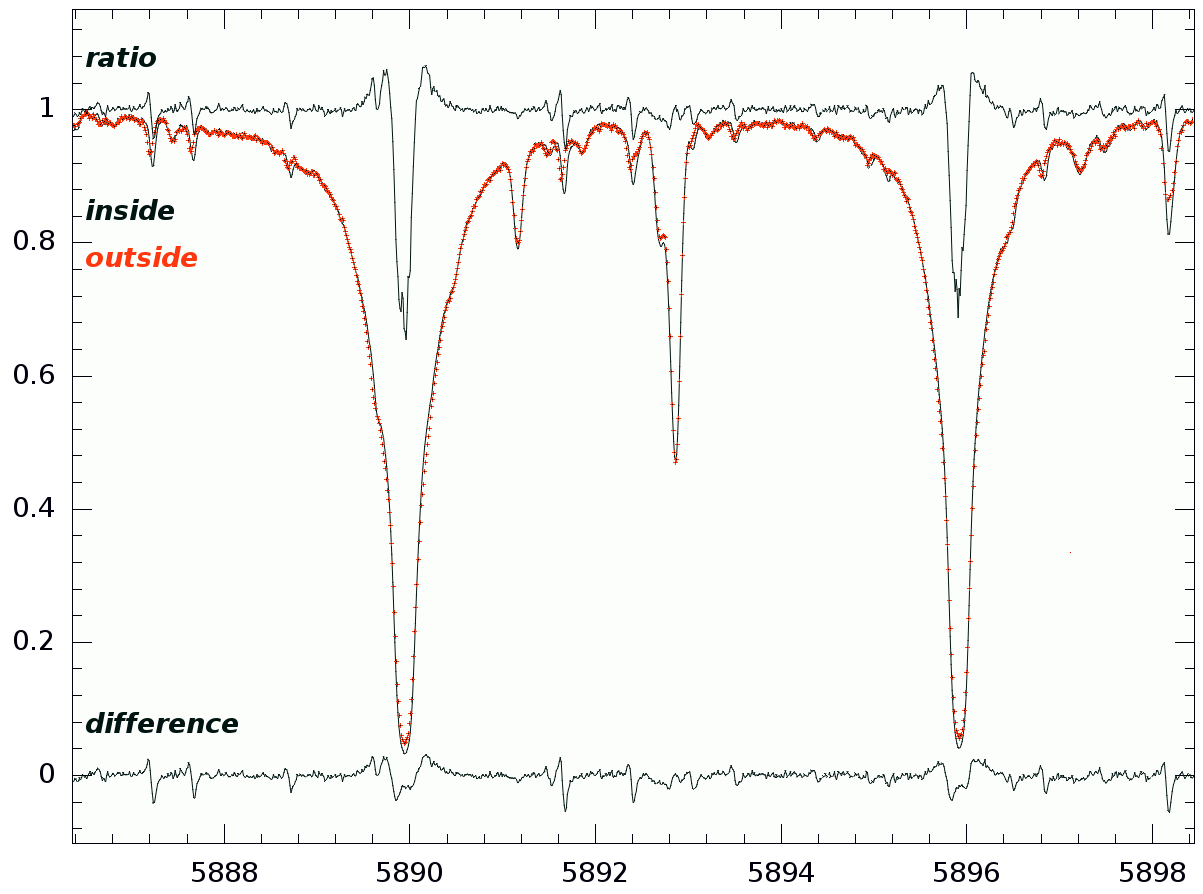}
\includegraphics[angle=0,width=42mm,clip]{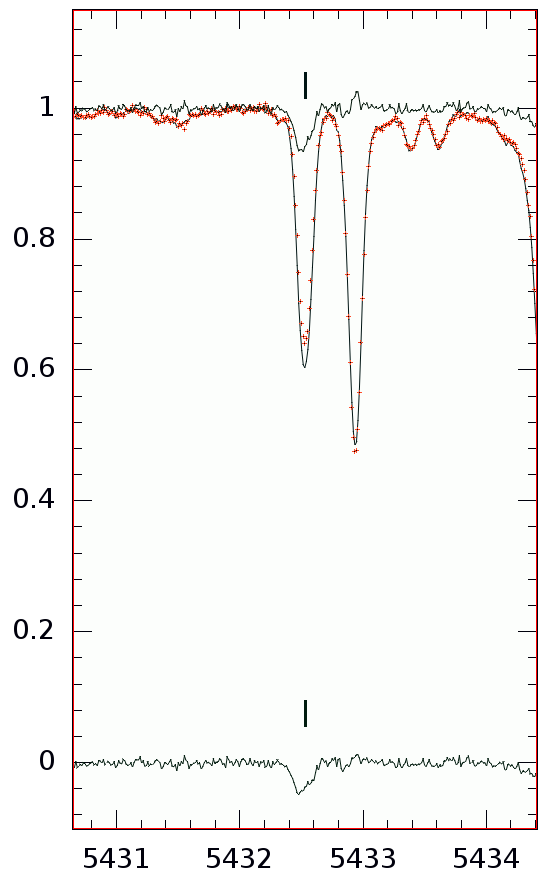}
\includegraphics[angle=0,width=42mm,clip]{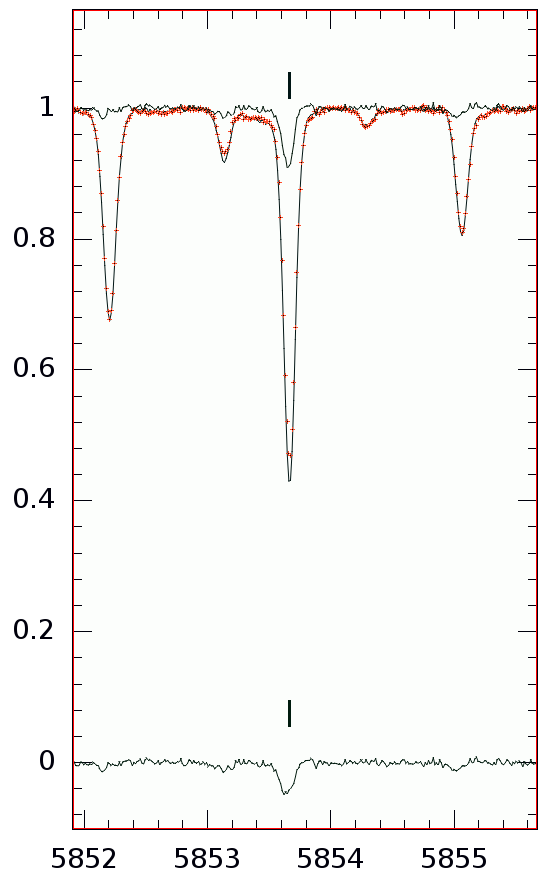}
\caption{Comparison of difference and ratio spectra at deep penumbral phase during ingress. \emph{a.} the Na\,{\sc i} D lines at 5890\,\AA\ and 5896\,\AA , \emph{b.} the Mn\,{\sc i}  line at 5432.5\,\AA\ (with a close Fe\,{\sc i} line at 5433\,\AA\ that does not show excess absorption) and, \emph{c.}, the Ba\,{\sc ii} line at 5853.7\,\AA. The two middle spectra in each panel are the observed spectrum outside eclipse (red pluses) and the Ring-corrected spectrum inside eclipse (black line). The spectrum at unit intensity is the ratio, and the spectrum at zero intensity the difference, always with respect to the observed lunar spectrum outside eclipse. All eclipse spectra are from UT\,3:50 during ingress. The $x$-axes are wavelength in \AA ngstroem. }
 \label{Fn}
\end{figure*}

The other effect that impacts on the transmission line depths is the solar center-to-limb variation (CLV). This is because the Sun becomes more and more a crescent before it completely disappears behind Earth (see Fig.~\ref{Fmaps}). Therefore, during the deep penumbral phases, we receive light only from regions near the solar limb. Most (optically thin) spectral lines tend to be strengthened toward the limb but some are not (see, e.g., Takeda \& Ueno \cite{tak:uen}, Allende-Prieto et al. \cite{all}). For the vast majority of lines the CLV effect works thus opposite to the Ring effect. Only singly-ionized lines with a large (lower) excitation potential show a decrease toward the limb or at least remain flat.

\begin{figure*}
\center
\includegraphics[angle=0,width=\textwidth,clip]{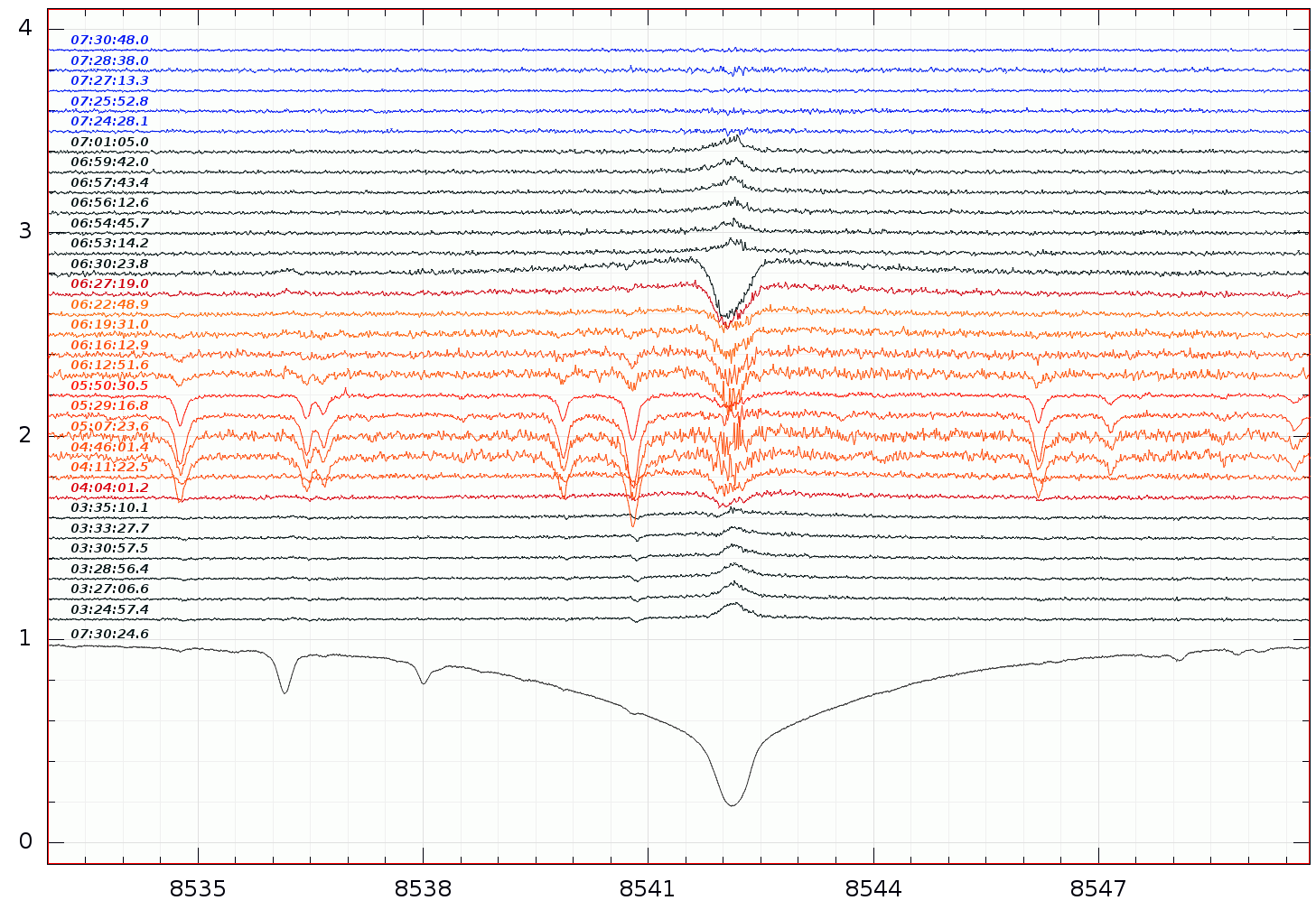}
\caption{Ratio spectra during eclipse with respect to out-of-eclipse for the Ca\,{\sc ii} IRT $\lambda$8542-\AA\ line. The bottom spectrum is the average out-of-eclipse spectrum from around UT\,7:30 used for the ratio. All three calcium IRT lines show excess absorption in the line core during eclipse. The excess absorption appears during the deepest penumbral phase at ingress at around UT\,3:35 and disappears at around UT\,6:50 shortly after the deepest penumbral phase at egress. It appeared twice as strong at egress than at ingress. This is likely only because the comparable egress phase was not covered with CD\,VI. The red-colored spectra in the figure indicate the umbral phases, the black colored the penumbral, and the blue colored the out-of-eclipse phases (top five spectra). The $x$-axis is wavelength in \AA ngstroem. We also note that time increases from bottom to top but is not equidistant.}
 \label{Firt}
\end{figure*}

In this paper, we correct for both effects by multiplying each CCD pixel with a constant. Within the wavelength range of a PEPSI cross disperser, we assume this constant to be the same for all pixels. We note that without a correction, strong solar absorption lines appeared typically in emission in the difference spectra (inside-minus-outside eclipse), that is the solar lines were generally weaker during eclipse than outside eclipse as expected. We select the strongest optically-thin lines in our spectra (typically Fe\,{\sc i} and Ca\,{\sc i} lines) and iteratively minimize their peak excess emission strengths in the difference spectra in order to find the best-fit constant per CD and per eclipse phase. The fit itself is based on a least-square minimization of the cross-correlation function between the in-eclipse spectrum and the out-of-eclipse spectrum for the respective wavelength regions. The Fraunhofer lines and their wings, as well as regions with strong telluric water-vapor or O$_2$ contamination, were excluded from the minimization. Figure~\ref{Fring} is a plot of these correction values as a function of time and for all CDs. We note that during the shallow penumbral eclipse phases the CLV effect dominates over the Ring effect and the corrections become smaller than unity. Values are between --2\% and +4.5\%. The remaining residual emission in the cores of, for example, the very strong Ca\,{\sc ii} IRT lines shown later in Fig.~\ref{Firt}, indicates that this correction is not fully satisfactory for the Fraunhofer lines and thus for the  correction of the actual Ring effect. Although one could easily increase the constant for the cores of the Fraunhofer lines, thereby removing the residual pseudo line-core emission during penumbral eclipse, there are physical limits up to which one can apply such a simple correction (see Arnold et al. \cite{luc2} for the Ring correction of Na\,D). Our approach of using the strongest non-resonance lines for the determination of a combined correction factor is thus a conservative approach with the disadvantage of only approximatively correcting the Ring effect. The advantage is that this minimal correction can only dilute the true excess absorption from Earth's atmosphere rather than enhance it, which makes a detection of excess absorption more safe but less significant.

\subsubsection{Penumbral spectra}

CDs~II, IV, and V (4264--4800, 5441--6278, and 6278--7419\,\AA , respectively) were additionally employed during some of the penumbral phases. These spectra are also water-vapor corrected with our time-dependent spline fit like for CD\,VI but the correction appears less precise due to the coarser sampling in time. CD\,V spectra were available only at egress (see the notes in Table~\ref{T1}).

We confirm residual Na\,{\sc i} D$_1$ and D$_2$ absorption in our corrected penumbral ratio and difference spectra in CD\,IV. Figure~\ref{Fn}a plots the spectrum from UT\,03:50 during deep penumbral ingress. In the Appendix in Fig.~\ref{F3App}, we show the full set of Na-D ratio spectra including phases from the egress. The excess absorption depth is highly significant from both methods with 14$\sigma$ from the difference spectra and a formal 60$\sigma$ from the ratio spectra, where $\sigma$ is determined by the ratio of the line depth to the residuals from a 100-pixel continuum range on both sides of the line. The spectrum's Ring-correction at UT\,3:50 was 2.2\%. The residual Na~D lines appear with broad pseudo emission wings left over from the simplified Ring correction (resembling the profiles from HARPS data in Arnold et al. \cite{luc2}). Our difference and ratio spectra reveal RV shifts of the water-vapor lines between outside and inside eclipse spectra and with respect to the spline fit to CD\,VI lines. However, these lines are easily recognizable because their residual profiles change sign with respect to the continuum and thus are easy to spot and do not impact the detection (see, e.g., the sharp wiggles in the difference spectrum in Fig.~\ref{Fn}a). The residual line widths of Na\,{\sc i}~D at continuum level are 0.38\,\AA\ and 0.31\,\AA\ for D$_2$ and D$_1$ and from both methods, respectively. Their average respective equivalent widths are 60\,m\AA\ and 50\,m\AA\ from the ratio spectra.

The Ca\,{\sc ii} IRT lines (8498, 8542, and 8662\,\AA) and K\,{\sc i} 7699\,\AA\ are also reconstructed with excess line-core absorption during the deep penumbral phases. The IRT absorption (Fig.~\ref{Firt}) appears again superimposed on pseudo emission line wings like for sodium, in particular at egress. We believe that these emission line wings are caused due to the same simplified Ring correction as for the Na\,{\sc i} D lines. The IRT line depths of the excess absorption is very significant though, even from the difference spectra; 8$\sigma$, 13$\sigma$, and 14$\sigma$ for the three IRT lines, respectively. Their absorption widths at continuum level are 0.46, 0.70, and 0.65\,\AA\ for the three lines, respectively, with an uncertainty of no more than $\pm$0.05\,\AA . Their respective residual equivalent widths are 20, 98, and 80\,m\AA\ during the one outstanding deep penumbral egress phase at UT\,6:30. We note that the exactly corresponding phase during ingress was not covered with CD\,VI (but with CD\,IV containing the Na\,D lines) and thus causes the impression of extra strong excess absorption be present only during egress. However, the Na\,D lines show a similar excess enhancement during the same deep penumbral phase at ingress. It is thus likely that the excess absorption occurs symmetrically mirrored along the eclipse as expected if from transmission through Earth's atmosphere.

The other significant detection during penumbral phase is potassium in form of the K\,{\sc i} 7699\,\AA\ line with even 17$\sigma$ in the difference spectrum at UT\,6:30, that is at the deepest penumbral phase at egress (Fig.~\ref{FK1}). The excess absorption varied by approximately a factor five and remained visible at all phases of the eclipse. Its maximum line width at continuum level was 0.29\,\AA\ (full width half maximum (FWHM) of 0.16\,\AA ). The residual equivalent width was 35\,m\AA\ at UT\,6:30 but only $\approx$7\,m\AA\ at UT\,5:50.

A detailed inspection of all residual penumbral spectra reveals the existence of two other spectral elements beyond our initial sodium, calcium and potassium detection. Manganese Mn\,{\sc i} lines that are broadened by hyper-fine-structure transitions in the solar spectrum remained as excess absorption lines in all residual spectra, for example, at $\lambda$5394.7 (8$\sigma$), 5407.3 (9$\sigma$), 5432.5 (11$\sigma$; Fig.~\ref{Fn}b), 5470.6 (6$\sigma$), 5516.8 (7$\sigma$), and 5537.8\,\AA\ (5$\sigma$). Their line widths at continuum level are on average 0.27\,\AA , their equivalent widths $\approx$10\,m\AA . We note again that line widths and standard deviations are measured from the difference spectra, equivalent widths from the ratio spectra, and that the absorption line depth is given always relative to the out-of-eclipse spectrum. Because we still see a few solar lines with excess emission (some strong Si\,{\sc i} and Fe\,{\sc i} lines) even after Ring and CLV correction, that is they appear stronger inside eclipse than outside eclipse, we adapt their residual strengths as our true detection limit. From five such lines within CD\,IV during UT\,3:50 and 6:30, we obtain 7$\sigma$ as a conservative threshold for a transmission detection. No significant excess absorption is seen neither in the otherwise very strong Mg\,{\sc i} triplet around 5170\,\AA\ in CD\,III at UT\,6:22 (Tycho then still in umbral eclipse) nor in the single Mg\,{\sc i} line at 5528.3\,\AA\ in CD\,IV at UT\,3:50 at ingress and 6:34 at egress (or in the Mg\,{\sc i} lines at $\lambda$8806.8 or 8736.0\,\AA\ within CD\,VI). Residual lines are also not seen, for example, from lithium at 6708\,\AA\ or from H$\alpha$ or H$\beta$, as expected. We also note that the atomic oxygen O\,{\sc i} triplet at 7774\,\AA\ nor the forbidden [O\,{\sc i}] 6300\,\AA\ or 8446\,\AA\ lines appear detectable in the ratio or difference spectra.

Quite surprising is a clear detection of excess absorption at the wavelengths of all barium lines visible in the optical solar spectrum, that is the barium red triplet Ba\,{\sc ii} 5853.7\,\AA , 6141.7\,\AA\ (blended with an iron line in the solar spectrum) and 6496.9\,\AA . The Ba\,{\sc ii} 5853.7\,\AA\ detection is even at 12$\sigma$ for the difference spectrum and has a line width at continuum of 0.20\,\AA\ and an equivalent width of 12\,m\AA\ (Fig.~\ref{Fn}c). The wavelength region that contains the very strong Ba\,{\sc ii} 4554\,\AA\ line was covered only with CD\,II. In the Sun this line is so strong that its core has a residual intensity of just 5\% of the continuum. Even our out-of-eclipse lunar spectrum reaches a S/N of just $\approx$40 in its line core, and the in-eclipse spectra only 17. This is insufficient for a meaningful ratio or difference spectrum.

Singly-ionized species usually can only be present in the Earth's ionosphere where the density is low but collision rates still sufficiently high. Large-scale geomagnetic storms and aurorae may cause anomalies to the ionospheric abundance and add extra transmission absorptions from Fe, Mg, or Si, for example (Roth \cite{roth}). It has been suggested that the amount of sodium in the upper Earth atmosphere can be temporarily increased by meteoritic deposition (Moussaoui et al. \cite{moussa}). At this point we can not provide an explanation why we detect atomic Na, Ca, K, Mn and Ba but not Mg and O.

\subsubsection{Umbral spectra}

Our observations sample the umbral eclipse practically only with CD\,VI for the wavelength range 7419–-9067\,\AA . The simultaneous blue-arm CD\,III for the wavelength range 4800–-5441\,\AA\ did not reach sufficient S/N during mid umbral eclipse (see Table~\ref{T1}) and is not used for analysis for these eclipse phases.

\begin{figure}
\center
\includegraphics[angle=0,width=85mm,clip]{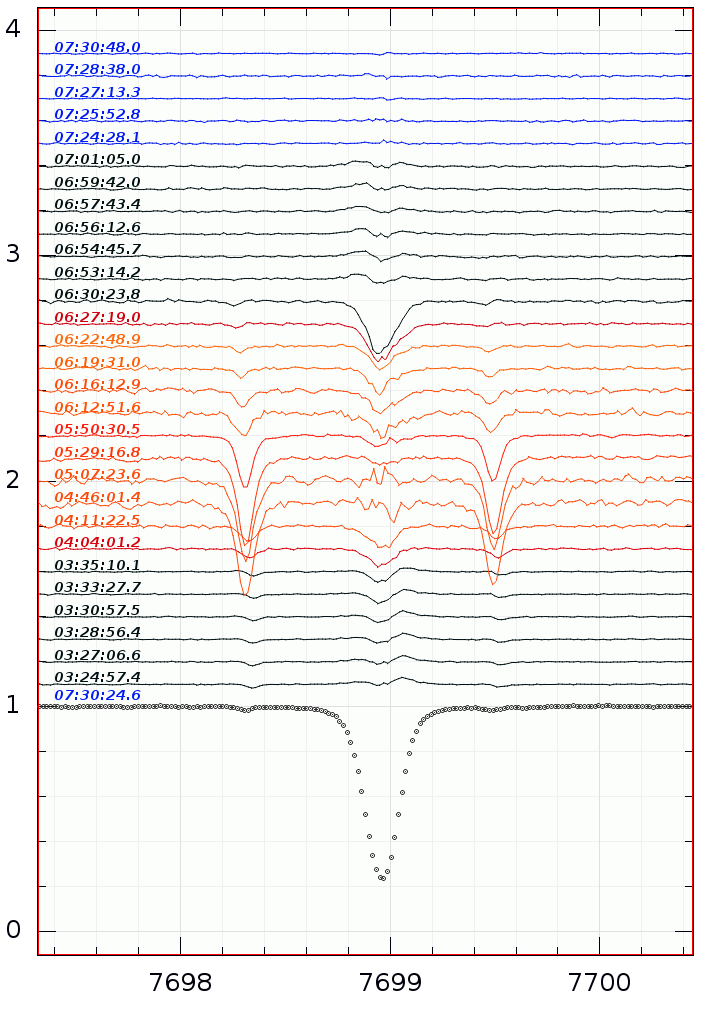}
\caption{Same as Fig.~9, but for the K\,{\sc i} $\lambda$7699-\AA\ line. Otherwise as in Fig.~\ref{Firt}. We note that the excess absorption in the line core is probably seen at all eclipse phases maybe with the exception of the exact mid eclipse at around UT\,5:07 (which is the noisiest spectrum of all though). The two tiny lines symmetric around K\,{\sc i}, at $\lambda$7698.304\,\AA\ and 7699.486\,\AA , are two water-vapor lines that became dramatically stronger during umbral eclipse. The out-of-eclipse spectrum is now plotted with dots to additionally show the CCD pixel sampling of PEPSI in its $R$=130\,000 mode. }
 \label{FK1}
\end{figure}

Figure~\ref{Firt} plots the entire time series of ratio spectra for the Ca\,{\sc ii} IRT $\lambda$8542-\AA\ line while Fig.~\ref{FK1} is the same for the potassium line at $\lambda$7699\,\AA. The main difference between the two lines is that the K\,{\sc i} line is more like an optically-thin line and does not have comparable  resonance wings. It thus appears without the uncertainties that the Ring correction causes for the line wings compared to the line core. The K\,{\sc i}-line Ring correction is thus more appropriate and results in a clearer absorption profile in the ratio spectra as compared to the Fraunhofer lines. The K-line width at continuum level is 0.33\,\AA\ at phase UT\,6:30. From the time series in Fig.~\ref{FK1}, we see that the excess line-core absorption remained throughout the eclipse, maybe with the exception of the exact mid-eclipse phase of exposure UT\,5:07 where the S/N per pixel is down to 80 (compared to 600--700 during penumbral eclipse). The line strength increases during ingress and decreases during egress, as expected. The Ca\,{\sc ii} IRT lines are also reconstructed with residual line-core absorption during umbral phases. Their residual line widths at continuum level are around 0.4--0.5\,\AA , and thus on average narrower than during the deep penumbral phases. The enhanced wings seen in residual Ca\,{\sc ii} at the two deep penumbral phases UT\,6:27 and UT\,6:30 of up to 7\,\AA\ are a remnant from the solar photospheric spectrum improperly removed, indicating that the single-constant Ring correction is indeed only a very rough approximation.

\begin{figure*}
{\bf a.} \hspace{71mm} {\bf b.}\\
\includegraphics[angle=0,width=69mm,clip]{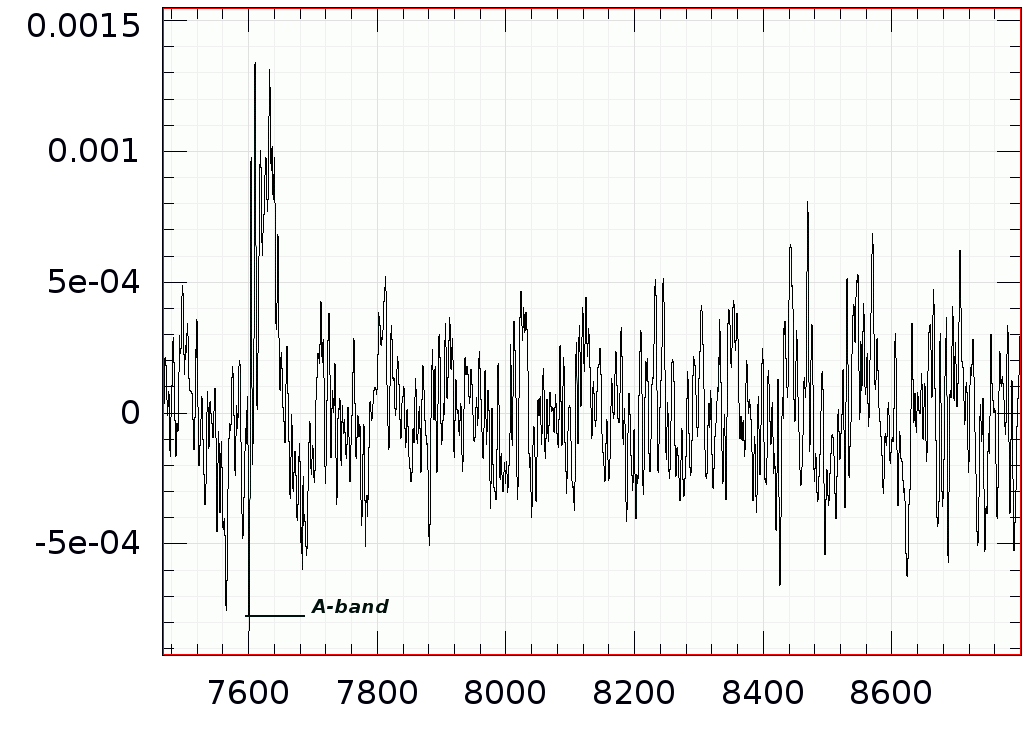}\hspace{5mm}
\includegraphics[angle=0,width=107mm,clip]{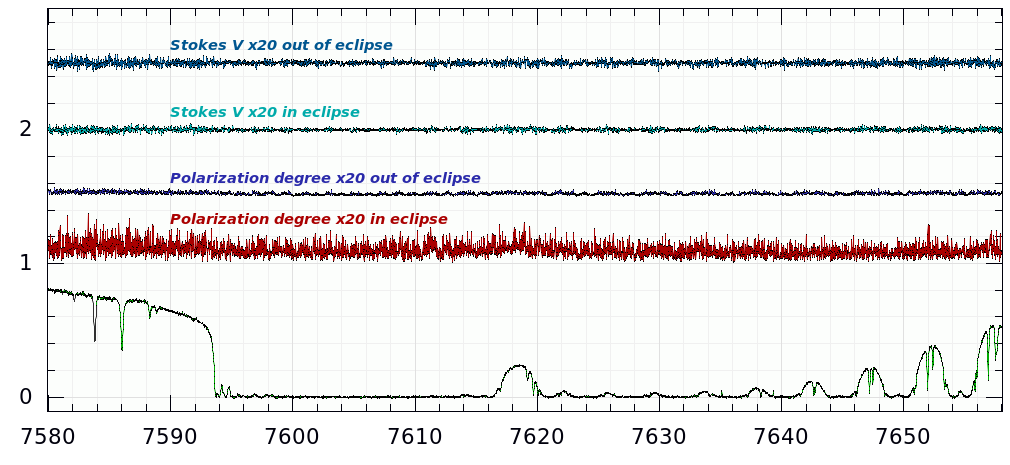}
\caption{\emph{a.} Degree of continuum polarization of the O$_2$ A-band during umbral eclipse. The $x$-axis is wavelength in \AA ngstroem. Shown is $Q/I$ from a convolution of the spectra with a Gaussian of 300\,\kms\ to prevent divisions by zero. The horizontal bar indicates the wavelength span of the major A-band lines. \emph{b.} Degree of line polarization around the pseudo emission line at 7618\,\AA\ outside and inside eclipse. The two top spectra are Stokes-V, the two middle spectra are the degree of polarization built from Q and U, the bottom spectrum is Stokes-I for comparison and shows the residual O$_2$ pseudo emission at 7618\,\AA . All polarized spectra are arbitrarily offset and enlarged by a factor of 20 compared to Stokes-I.}
 \label{F11}
\end{figure*}

There is the clear detection of increased H$_2$O absorption during mid-umbral eclipse (see Fig.~\ref{F2} for the full CD\,VI wavelength range) verifying the earlier detections by Pall\'e et al.~(\cite{palle}) and Vidal-Madjar et al. (\cite{vidal}). Our spectra show a maximum H$_2$O line strength at UT\,4:46 until 5:50, well within umbral eclipse, opposite to the atomic absorbers Ca, Na, and K which show their maximum during deep penumbral eclipse.  Because water signatures should be confined to low altitudes on Earth the increased water-vapor absorption during umbral eclipse is expected. Garc\'ia Munoz et al.~(\cite{garcia}) had shown that refraction dominates the umbral spectrum and that sunlight reaches heights in the Earth's atmosphere of 12--14\,km during a mid-eclipse configuration while Arnold et al. (\cite{luc2}) confirmed water-vapor lines from altitudes as low as 20\,km.

\subsection{Searching for continuum and line polarization}\label{S34}

Our final step in the analysis in this paper is to recover the continuum polarization of the Earth's transmission spectrum  and then search for line polarization. Because of the faintness of the Moon during totality, we have only one QUV spectrum during umbral eclipse. It was taken around mid eclipse phases when the angle formed by the Sun, the Moon, and the Earth is almost exactly 0\degr . Common thinking would suggest that continuum linear polarization (LP) be at a maximum if from coherent scattering vertical to the Earth's surface. However, in its current setup PEPSI is not prepared to measure continuum LP but only line LP. Continuum LP from PEPSI would require the use of a $\lambda/2$ retarder, which had not been foreseen because LP in PEPSI is extracted by rotating the Foster beam-splitting prism without a retarder (to bypass cross talk). Animated by the referee, we tried to recover the continuum signal from just the beam splitter. It simply means that we build the LP only from I$\pm$Q but not from I$\mp$Q (basically absolute polarimetry versus differential polarimetry). For such a trivial approach we keep all unwanted seeing and instrumental polarization effects from the Foster prism in the data and any results are only moderately precise at best. We then define the degree of polarization to be simply $P=Q/I$ while missing out on any slopes or zero-point differences between the ordinary and the extra-ordinary beam versus wavelength. However, the problem of division through zero remains for the O$_2$ A-band wavelengths when building $P$. We mitigate this by broadening  our spectra to 300\,\kms\ (i.e., FWHM of 7.6\,\AA\ at 7600\,\AA ) before division by $I$. Figure~\ref{F11}a shows the umbral $P$-spectrum at UT\,04:04 just after Tycho went into umbral eclipse. The O$_2$ A-band appears as the only significant (6.3$\sigma$) signal with a 0.12\%\ degree of continuum polarization. This is comparable to the 0.6\,\%\ residual A-band polarization seen by Sterzik et al. (\cite{sterz}) from VLT/FORS Earthshine observations with a 30-\AA\ FWHM in 2011. It is weaker though than what Takahashi et al. (\cite{tak2}) had obtained during the eclipse in 2014, and also below the expected peak from the coarse model spectrum with FWHM=10\,\AA\ in Emde et al. (\cite{emde}).

For the search for line polarization, we sum the LP spectra to represent the degree of polarization now defined as the square root of the squared sum of Stokes Q and U; $P=(Q^2+U^2)^{1/2}/I_c$. Figure~\ref{F11}b compares outside eclipse and inside eclipse $P(\lambda)$ spectra for the wavelength region of the O$_2$ pseudo emission around 7618\,\AA . No line polarization is detected neither outside nor inside eclipse. The tiny bump at the residual O$_2$ emission at 7618\,\AA\ is below 1$\sigma$ and is not considered real. Circular line polarization could be introduced by the Zeeman-effect when sunlight passes through the Earth's magnetic field. If only the transverse Zeeman-effect would be at work and we assume a radial field density of the Earth's atmospheric magnetic field of 1\,G, we would expect LP with a signal strength of only 10$^{-4}$-10$^{-5}$. Then, our Stokes-QU data would not have adequate S/N to detect it. If also a longitudinal effect contributes or even dominates then it adds a Stokes-V signal possibly as large as $10^{-3}$-$10^{-4}$. In Stokes-I the line splitting would be of order 1\,\ms\ and possibly detectable for PEPSI (see Strassmeier et al. \cite{sun}). We did not detect either of these proxies in the present spectra. Sterzik et al. (\cite{sterz}) also did not detect circular polarization in the spectrum of the Earthshine. Additionally, linear line polarization could be introduced into the transmission spectrum by coherent scattering at electrons in the Earth's ionosphere, just like the Hanle-effect on the Sun, and then adds to the total polarization signal. Fauchez et al. (\cite{fau}) showed that gaseous absorption bands like the O$_2$ bands not only show up in flux spectra of light reflected by (exo)planets but also appear in polarization spectra. However, our QUV data do not show any polarization above 0.2--0.4\%. Any Earth-induced line polarization during transmission remained thus undetected. We note that a null detection outside eclipse, that is in Stokes spectra from the Sun as a star, is expected because of the low activity of the Sun in early 2019 and the general disk-integrated magnetic flux cancelation.

A remaining issue is depolarization by the reflection on the Moon. The dielectric state of the reflecting and scattering lunar surface introduces a certain preference in polarization angle, which is wavelength dependent in the sense longer wavelengths become more depolarized than shorter wavelengths (Dollfus \cite{doll}). Moreover, the lunar surface shows high spatial frequency polarimetric structures not resolved by seeing-limited data (e.g., Shkuratov et al. \cite{shk}). On larger scales, the Moon's dark mare are more polarized than the bright highland terrains (Coyne \& Pellicori \cite{coyne}). Therefore, observations suffer additionally from different levels of depolarization due to the back-scattering from the lunar soil. On average, we must assume a depolarization of between a factor two to three. The true polarization degree of the reflected Earth spectrum is thus not well known (see Emde et al. \cite{emde}).

\section{Summary and conclusions}\label{S4}

We present high-resolution time-series spectroscopy and polarimetry of the January 21st, 2019 total lunar eclipse. The exposure meter of our spectrograph was used as a photon counting photometer in white light and recorded a 10.75 magnitude drop of brightness during totality. The projected fiber diameter of the spectrograph provided light from 1.77 squared arc seconds of the Moon's surface to the spectrograph. It translates to an approximate $V$ brightness of 13\fm5 during umbral eclipse. For a spectral resolution of 130\,000, and in particular for Stokes-IQUV polarimetry, this would be otherwise dubbed a faint star.

The intensity spectra during umbral eclipse are dominated by strong O$_2$ and H$_2$O atmospheric flux absorption otherwise not seen as strong during penumbral eclipse or outside eclipse. The presence of the pre-dominantly low-altitude H$_2$O absorption during umbral eclipse suggests significant water vapor at heights between $\approx$12--20\,km above ground as already anticipated by Arnold et al. (\cite{luc2}). When removing the solar spectrum from the umbral observations, as well as the telluric absorption from the second transmission through Earth's atmosphere after reflectance off the Moon, we are left with an approximate transmission spectrum of Earth as an exoplanet. At this point we reemphasize that our high-resolution spectra are corrected for telluric and stray-light contamination by using the spectra themselves rather than simultaneous and independent sky measurements.  Beside excess molecular absorption, our umbral spectra also show excess atomic absorption in the line cores of the singly-ionized Ca\,{\sc ii} infrared triplet at a 8--14$\sigma$ level as well as in the K\,{\sc i} line at 7699\,\AA\ at even 17$\sigma$. Recently, potassium was detected in the atmosphere of the hot Jupiter HD\,189733b with a 7$\sigma$ significance using high-resolution transmission spectra (Keles et al. \cite{keles}) as well as ionized calcium in Wasp-33b and Kelt-9b (Yan et al. \cite{yan2}).

From our deep penumbral spectra, we additionally identify excess absorption from the neutral Na~D lines (at 14$\sigma$ significance), several neutral Mn lines (5--11$\sigma$), and singly-ionized Ba (7--12$\sigma$). The line widths of these excess absorptions are between 0.33\,\AA\ (K\,{\sc i}) and 0.70\,\AA\ (Ca\,{\sc ii} IRT) and thus broader than the excess telluric H$_2$O absorptions of $\approx$0.23\,\AA\ at continuum level. The fact that the solar IRT line with the strongest wings (Ca\,{\sc ii} 8542\,\AA ) shows the broadest excess absorption, while the solar line without resonance wings (K\,{\sc i}) shows the most narrow excess absorption, suggests that the observed line widths are still contaminated by the solar spectrum rather than solely Earth induced. Together with the fact that we see a few solar Si\,{\sc i} and Fe\,{\sc i} lines remaining after our center-to-limb correction indicates a threshold of $\approx$7$\sigma$ in the difference spectra as our true significance limit for the detection of excess absorption. The Mn lines are special because all of them are hyper-fine-structure lines. A disk-resolved spectrum in Fig.~2 in Takeda \& Ueno (\cite{tak:uen}) showed the solar Mn\,{\sc i} 5432.5\,\AA\ line dramatically stronger (in the core and the wings) at the limb than in the disk center compared to any of the other lines in this figure. However, no one did explicitly measure Mn lines for CLV so far. Because our CLV correction is from an ensemble of lines, and thus an average, it is then likely inappropriate for Mn lines. Therefore, we do not claim that its excess absorption is actually due to the transmission through Earth's atmosphere but rather suggest that it is due to an enhanced solar CLV effect.

The detection of singly-ionized barium is particularly interesting. Barium is never found free on Earth since it reacts with oxygen in the air, forming barium oxide (BaO). Barium enters the air during volcanic activity and man-made mining and refining processes and coal and oil combustion. Since the 1950s artificial barium injections with sounding rockets are used to probe  high-altitude winds and the Earth's magnetic and electric field between 150--250\,km height (see, e.g., F\"oppl et al. \cite{flop}). Naturally occurring barium is a mixture of seven stable isotopes, with $^{138}$Ba being the most dominant. Its ionization energy of 5.2\,eV being similar to Li, Na, or K. It exists in the solar photosphere mostly in singly-ionized form (Holweger \& M\"uller \cite{hol:mue}), and was studied in other stars as early as Burwell (\cite{bur}); see also Korotin et al. (\cite{koro}) for more recent measurements. An abundance relation of the Fe-peak element Mn and the s-process element Ba exists (e.g., Allen \& Porto de Mello \cite{all:por}) which may argue in favor of the reality of our detections of both elements together. However, because singly-ionized solar lines with a high excitation potential become usually weaker at the solar limb (Takeda \& Ueno \cite{tak:uen}), and because no studies exist for the Ba\,{\sc ii} lines, we do not claim in this paper that the Ba excess absorption is due to transmission through Earth's atmosphere but rather suspect that it is also due to an enhanced solar CLV effect. In any case, we caution the reader because we had no independent simultaneous sky spectra for removal but instead  relied on numerical sky removal using the actual observations.

Our spectra are also used to determine precise radial velocities from cross correlation of each intensity spectrum in the time series with the IAG FTS spectrum of the Sun. It returns the radial velocities of the Sun during an Earth eclipse as seen from the Moon and thus trace the well-known Rossiter-McLaughlin effect due to alternately masking half of the solar disk with its respective opposite directions of rotation. Our velocities show a larger RV amplitude from blue-wavelength spectra at $\approx$4500\,\AA\ than from red-wavelength spectra at $\approx$8200\,\AA\ by 47\,\ms , as already found by Yan et al. (\cite{yan}). This is due to stronger Rayleigh scattering in the blue which makes the atmosphere more opaque and the atmospheric effective thickness larger.

We did not detect a signal in the line LP nor the line CP for the wavelength range 7419--9067\,\AA . Such a signal could be evident if either the Zeeman effect or selective coherent scattering would act on the photons along their path through the Earth's atmosphere and magnetosphere. From solar observations, we would expect a signal of $\Delta P/P$$\approx$10$^{-4}$ with respect to the continuum, or even less, and thus require S/N of at least 10$^4$. Our best in-eclipse polarimetric spectrum still provides S/N$\approx$300 per pixel but could not reveal such low signals. Therefore, we can confine any line polarization to being smaller than 0.2\%. This line polarization is not to be confused with the 2\%\ disk-integrated continuum polarization first seen by Coyne \& Pellicori (\cite{coyne}) which is explained by double scattering in the Earth's atmosphere. We extracted continuum polarization from our Foster beam splitter alone, that is without a $\lambda/2$ retarder, and obtained a 6.3$\sigma$ 0.12\%\ degree of continuum polarization in the O$_2$ A-band during umbral eclipse, significantly weaker than the 1--3\,\%\ seen by Takahashi et al. (\cite{tak}) for the lunar eclipse in 2014. We consider our continuum polarization very uncertain though.

All reduced PEPSI spectra can be downloaded from our instrument web page\footnote{PEPSI data: https://pepsi.aip.de}. Integral-light spectra may also be retrieved from CDS/Strasbourg.

\acknowledgements{We thank all AIP and LBTO engineers and technicians involved in PEPSI. PEPSI was made possible by funding through the State of Brandenburg (MWFK) and the German Federal Ministry of Education and Research (BMBF) through their Verbundforschung grants 05AL2BA1/3 and 05A08BAC. The German Rat Deutscher Sternwarten (RDS) is thanked for granting the LBT time for this program. An anonymous referee is deeply thanked for the constructive criticism and the many helpful comments which resulted in a substantially improved paper. FM conducted the work in this paper in the framework of the International Max-Planck Research School (IMPRS) for Solar System Science at the University of G\"ottingen (Volkswagen Foundation project grant number VWZN3020). LBT Corporation partners are the University of Arizona on behalf of the Arizona university system; Istituto Nazionale di Astrofisica, Italy; LBT Beteiligungsgesellschaft, Germany, representing the Max-Planck Society, the Leibniz-Institute for Astrophysics Potsdam (AIP), and Heidelberg University; the Ohio State University; and the Research Corporation, on behalf of the University of Notre Dame, University of Minnesota and University of Virginia. PEPSI home page is at https://pepsi.aip.de/. }

\appendix

\section{Telluric contribution to CD-VI spectra}

\begin{figure}[htb]
{\bf a.} \\
\includegraphics[angle=0,width=86mm,clip]{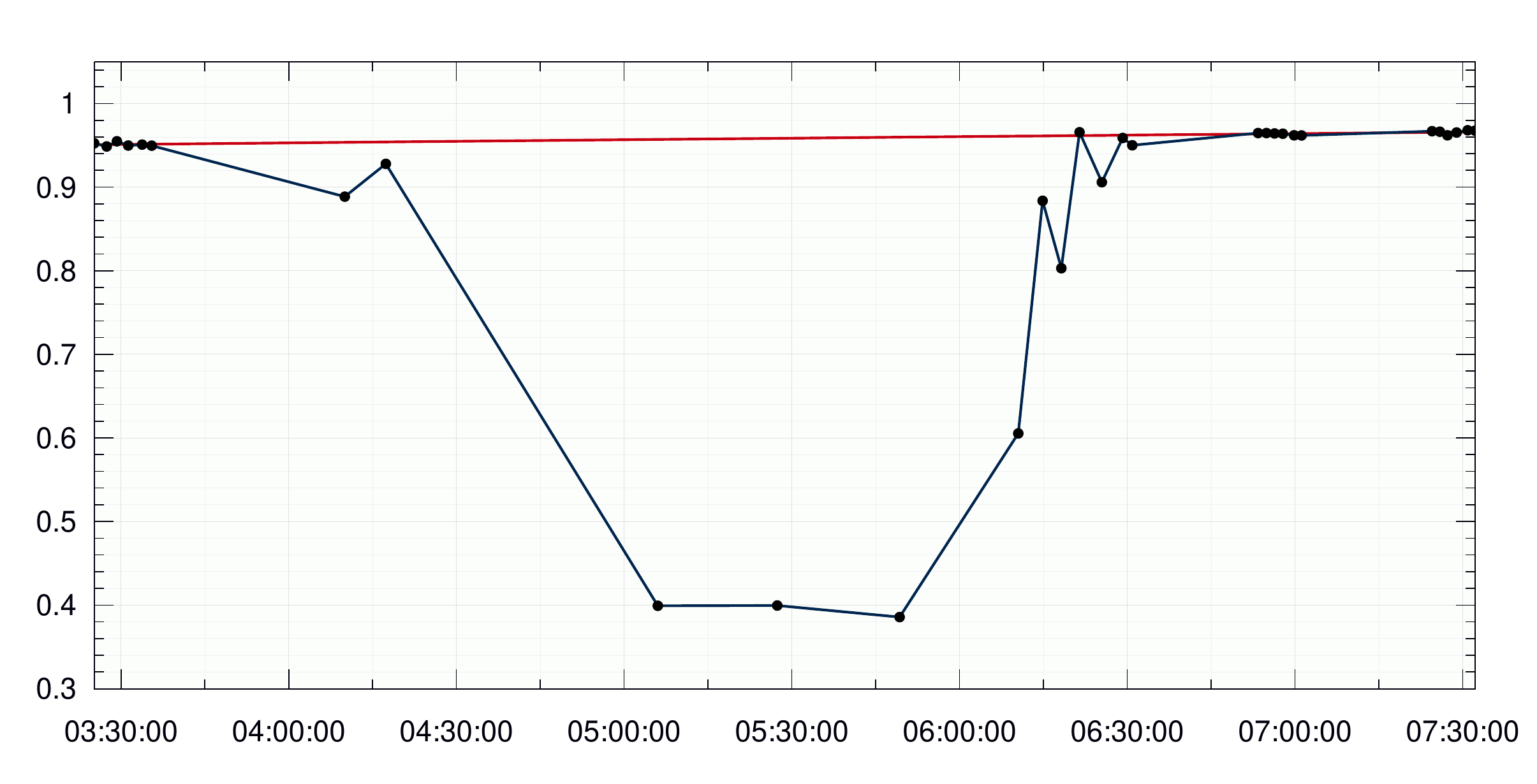}

{\bf b.}\\
\includegraphics[angle=0,width=86mm,clip]{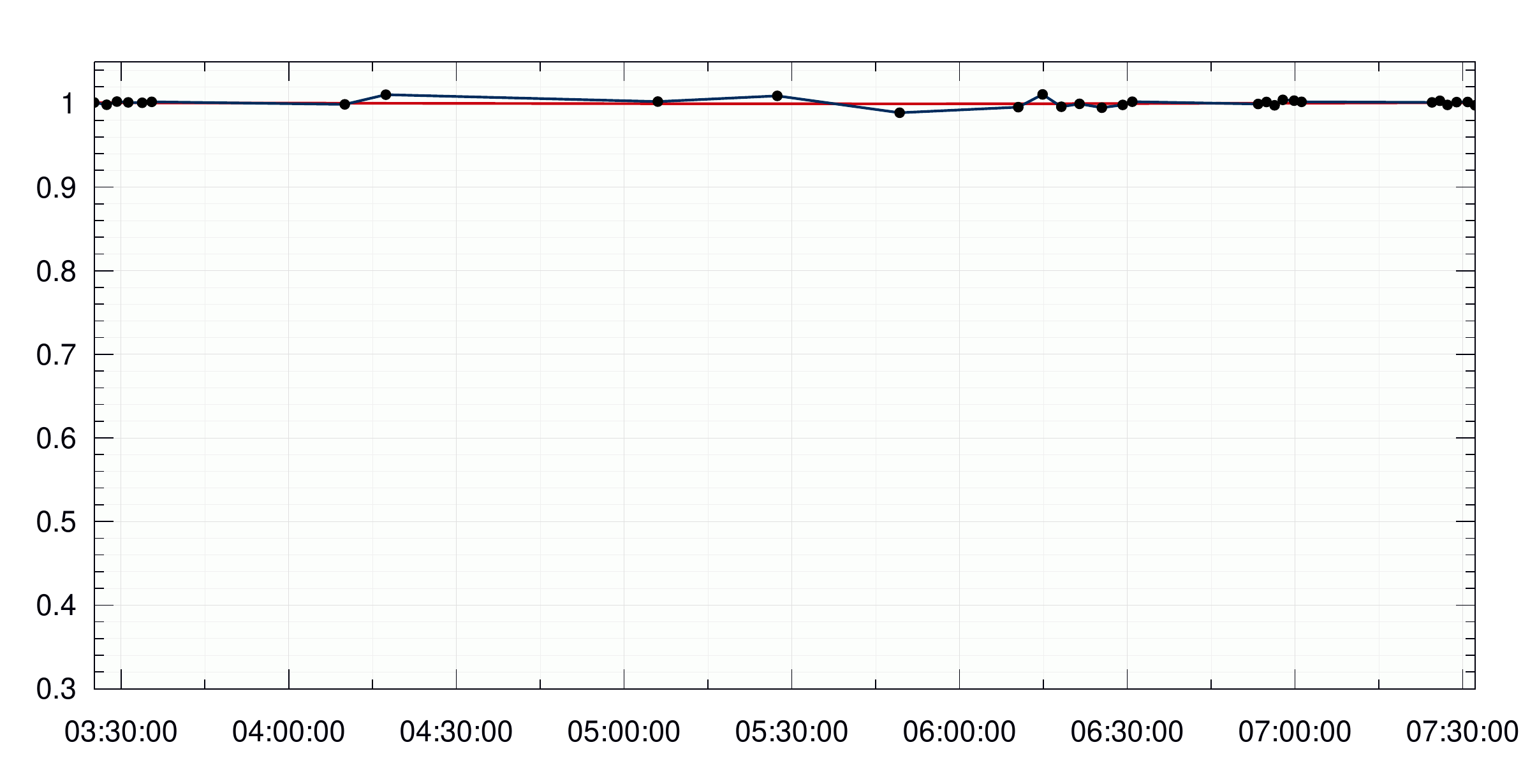}
\caption{Relative intensity of the telluric contribution to the CD-VI spectra at two wavelengths. Panel \emph{a.} at  7618\,\AA\ with a strong telluric line, \emph{b.} at 7585\,\AA\ without significant telluric contamination. The dots were derived from a spline fit to every pixel versus time. The top smooth line is the low-order spline fit that represents the air mass changes during our observations. $x$-axis is time in UT. }
 \label{F1App}
\end{figure}

\begin{figure*}
\includegraphics[angle=0,width=\textwidth,clip]{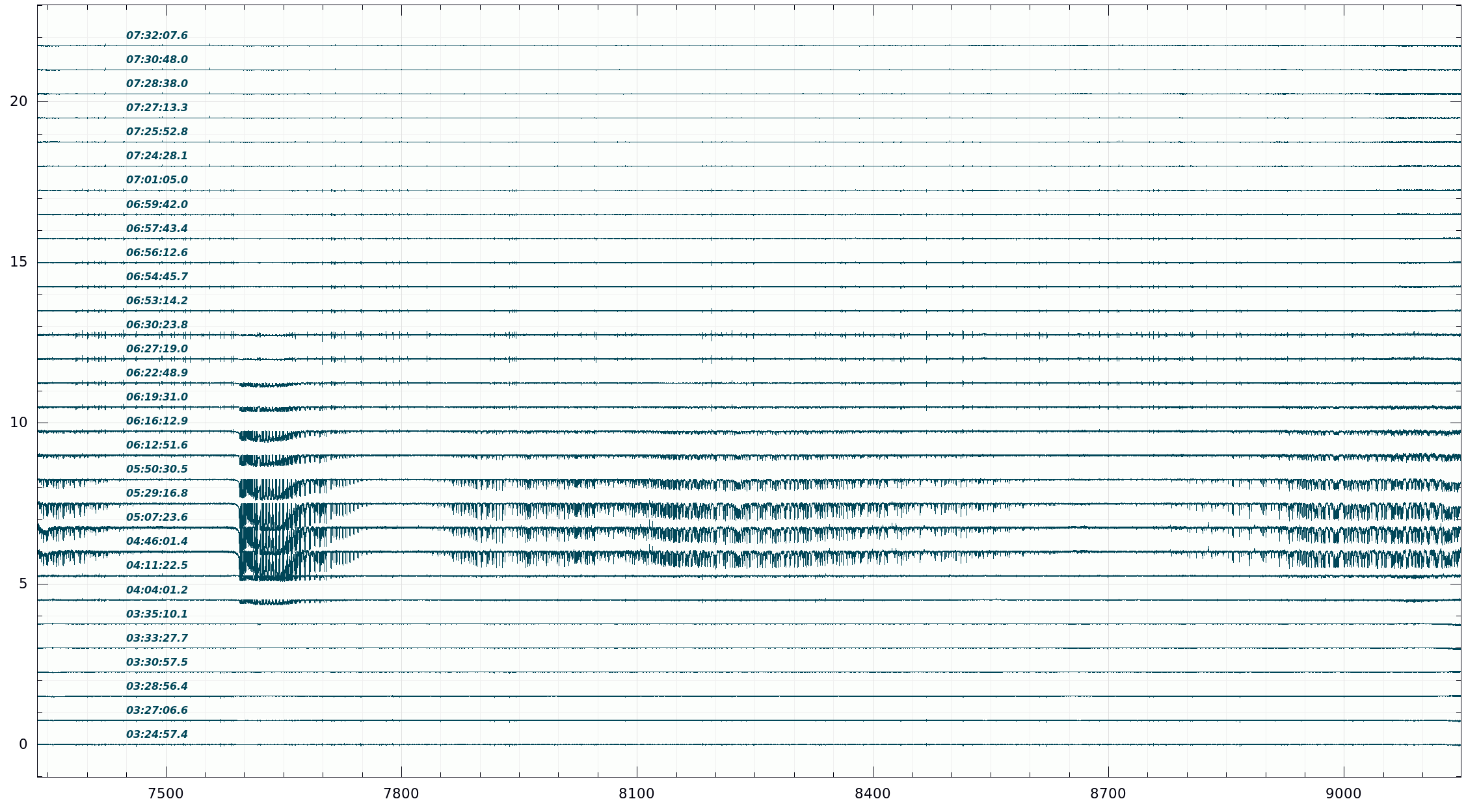}
\caption{Difference spectra after removal of both the solar spectrum and the telluric absorptions from the observed spectra. The remaining residuals are then solely due to the first transmission through Earth's atmosphere. Shown are all available CD-VI spectra. A vertical offset of 0.75 is applied for better visibility. Times on the left are start of exposure in UT. Wavelengths are in \AA ngstroem. }
 \label{F2App}
\end{figure*}

The spectra in CD-VI were re-binned into the same wavelength scale and placed into an image versus time. The intensities in each wavelength pixels were then fitted with a low degree smoothing spline versus time. The spline uses robust statistics to detect and mask-out any out-layers of the fit, so that the points which obey the normal distribution remain and are used for the fit. This is just the same as if we would make a spectrum continuum fit versus wavelength, but here is versus time. The fitted surface is then subtracted from the original image to form the residual transmission image of the lunar eclipse retaining the points with the maximal absorption and leaving unaffected wavelength pixels at zero. The wavelength scale used here is the laboratory wavelength not corrected for any radial velocities. The solar spectrum is changing its wavelength due to Earth rotation and the airmass is changing with height above horizon which both result in a tilt of solar lines in wavelength and a telluric line depth versus time. These two temporal variations are taken out with an appropriate smoothing spline of a degree not affecting the extraction of the transmission spectrum. The ``Transmission spectrum'' in Fig.~\ref{F6} is the weighted average of the residual spectrum in umbra. The ``Solar + Telluric spectrum'' is the average of the spline fit of the surface. Figure~\ref{F1App} shows the spline fit (red line) to the continuum points before and after eclipse at 7618\,\AA\ compared to region at 7585\,\AA\  not affected by the lunar eclipse. Figure~\ref{F2App} is the application of the fit to all CD-VI spectra as a function of time.

\begin{figure*}
\includegraphics[angle=0,width=\textwidth,clip]{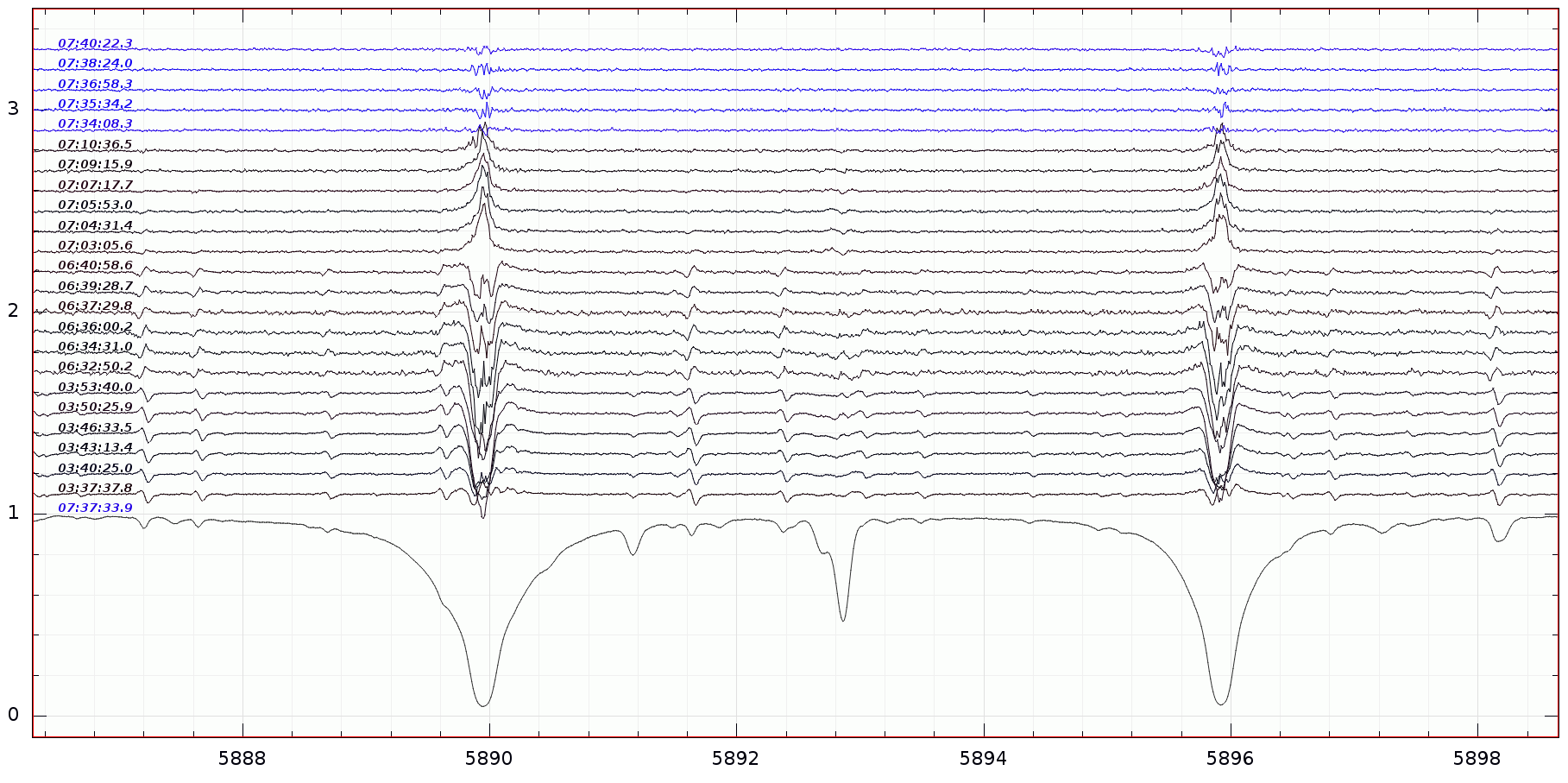}
\caption{Na\,D ratio spectra during penumbral eclipse (black) compared to out-of-eclipse (blue; top five spectra). Each spectrum is labeled with its UT at start of exposure. The bottom spectrum is a single out-of-eclipse spectrum. The deeper penumbral phases are reconstructed with excess Na~D absorption while the shallow penumbral phases (UT\,7:03--7:10) are not. We also note that the strong excess absorption during the deep penumbral ingress at UT\,3:50 could not have been seen with CD\,VI (which includes the O$_2$ A-band, the neutral K line, and the Ca IRT) because CD\,IV was used instead. The residual excess emissions during shallow penumbral eclipse are due to the improper Ring corrections for these phases (see text). Wavelengths are in \AA ngstroem and a vertical offset of 0.1 is applied. }
 \label{F3App}
\end{figure*}

\section{Observing log and radial velocities}

\begin{table*}[!tbh]
\caption{Observing log of all individual polarimetric CD-VI spectra}\label{T1-App}
\begin{tabular}{lllllll}
\hline\hline\noalign{\smallskip}
 UT start    & $t_{\rm exp}$ & $\Delta\lambda$ & Beam & Retarder     & Foster         & $<S/N>$ \\
 (hh:mm:ss)  & (mm:ss)   & (\AA )    &      & \multicolumn{2}{c}{(degrees)} & \\
\noalign{\smallskip}\hline\noalign{\smallskip}
 03:24:57.4 & 00:15.000  & 7419 - 9067 &  DXe & \dots &   0.0 & 386\\
 03:24:57.4 & 00:15.000  & 7419 - 9067 &  DXo & \dots &   0.0 & 349\\
 03:24:57.4 & 00:15.000  & 7419 - 9067 &  SXe & \dots &   0.0 & 320\\
 03:24:57.4 & 00:15.000  & 7419 - 9067 &  SXo & \dots &   0.0 & 338\\
 03:27:06.6 & 00:20.004  & 7419 - 9067 &  DXe & \dots &  90.0 & 364\\
 03:27:06.6 & 00:20.004  & 7419 - 9067 &  DXo & \dots &  90.0 & 328\\
 03:27:06.6 & 00:20.004  & 7419 - 9067 &  SXe & \dots &  90.0 & 299\\
 03:27:06.6 & 00:20.004  & 7419 - 9067 &  SXo & \dots &  90.0 & 318\\
 03:28:56.4 & 00:20.000  & 7419 - 9067 &  DXe & \dots &  45.0 & 392\\
 03:28:56.4 & 00:20.000  & 7419 - 9067 &  DXo & \dots &  45.0 & 354\\
 03:28:56.4 & 00:20.000  & 7419 - 9067 &  SXe & \dots &  45.0 & 326\\
 03:28:56.4 & 00:20.000  & 7419 - 9067 &  SXo & \dots &  45.0 & 347\\
 03:30:57.5 & 00:20.000  & 7419 - 9067 &  DXe & \dots & 135.0 & 371\\
 03:30:57.5 & 00:20.000  & 7419 - 9067 &  DXo & \dots & 135.0 & 336\\
 03:30:57.5 & 00:20.000  & 7419 - 9067 &  SXe & \dots & 135.0 & 308\\
 03:30:57.5 & 00:20.000  & 7419 - 9067 &  SXo & \dots & 135.0 & 328\\
 03:33:27.7 & 00:20.000  & 7419 - 9067 &  DXe & 45.0 &   0.0 & 344\\
 03:33:27.7 & 00:20.000  & 7419 - 9067 &  DXo & 45.0 &   0.0 & 306\\
 03:33:27.7 & 00:20.000  & 7419 - 9067 &  SXe & 45.0 &   0.0 & 279\\
 03:33:27.7 & 00:20.000  & 7419 - 9067 &  SXo & 45.0 &   0.0 & 299\\
 03:35:10.1 & 00:20.000  & 7419 - 9067 &  DXe &135.0 &   0.0 & 328\\
 03:35:10.1 & 00:20.000  & 7419 - 9067 &  DXo &135.0 &   0.0 & 286\\
 03:35:10.1 & 00:20.000  & 7419 - 9067 &  SXe &135.0 &   0.0 & 264\\
 03:35:10.1 & 00:20.000  & 7419 - 9067 &  SXo &135.0 &   0.0 & 281\\
 04:04:01.2 & 06:00.000  & 7419 - 9067 &  DXe & \dots &   0.0 & 182\\
 04:04:01.2 & 06:00.000  & 7419 - 9067 &  DXo & \dots &   0.0 & 153\\
 04:04:01.2 & 06:00.000  & 7419 - 9067 &  SXe & \dots &   0.0 & 136\\
 04:04:01.2 & 06:00.000  & 7419 - 9067 &  SXo & \dots &   0.0 & 149\\
 04:11:22.5 & 06:00.000  & 7419 - 9067 &  DXe & \dots &  90.0 & 115\\
 04:11:22.5 & 06:00.000  & 7419 - 9067 &  DXo & \dots &  90.0 &  97\\
 04:11:22.5 & 06:00.000  & 7419 - 9067 &  SXe & \dots &  90.0 &  82\\
 04:11:22.5 & 06:00.000  & 7419 - 9067 &  SXo & \dots &  90.0 &  91\\
 04:46:01.4 & 20:00.000  & 7419 - 9067 &  DXe & \dots &  45.0 &  55\\
 04:46:01.4 & 20:00.000  & 7419 - 9067 &  DXo & \dots &  45.0 &  45\\
 04:46:01.4 & 20:00.000  & 7419 - 9067 &  SXe & \dots &  45.0 &  35\\
 04:46:01.4 & 20:00.000  & 7419 - 9067 &  SXo & \dots &  45.0 &  39\\
 05:07:23.6 & 20:00.000  & 7419 - 9067 &  DXe & \dots & 135.0 &  46\\
 05:07:23.6 & 20:00.000  & 7419 - 9067 &  DXo & \dots & 135.0 &  38\\
 05:07:23.6 & 20:00.000  & 7419 - 9067 &  SXe & \dots & 135.0 &  28\\
 05:07:23.6 & 20:00.000  & 7419 - 9067 &  SXo & \dots & 135.0 &  32\\
 05:29:16.8 & 20:00.000  & 7419 - 9067 &  DXe & 45.0 &   0.0 &  97\\
 05:29:16.8 & 20:00.000  & 7419 - 9067 &  DXo & 45.0 &   0.0 &  82\\
 05:29:16.8 & 20:00.000  & 7419 - 9067 &  SXe & 45.0 &   0.0 &  64\\
 05:29:16.8 & 20:00.000  & 7419 - 9067 &  SXo & 45.0 &   0.0 &  72\\
 05:50:30.5 & 20:00.000  & 7419 - 9067 &  DXe &135.0 &   0.0 & 165\\
 05:50:30.5 & 20:00.000  & 7419 - 9067 &  DXo &135.0 &   0.0 & 145\\
 05:50:30.5 & 20:00.000  & 7419 - 9067 &  SXe &135.0 &   0.0 & 124\\
 05:50:30.5 & 20:00.000  & 7419 - 9067 &  SXo &135.0 &   0.0 & 138\\
 06:12:51.6 & 02:00.000  & 7419 - 9067 &  DXe & \dots &   0.0 &  60\\
 06:12:51.6 & 02:00.000  & 7419 - 9067 &  DXo & \dots &   0.0 &  51\\
 06:12:51.6 & 02:00.000  & 7419 - 9067 &  SXe & \dots &   0.0 &  44\\
 06:12:51.6 & 02:00.000  & 7419 - 9067 &  SXo & \dots &   0.0 &  49\\
 06:16:12.9 & 02:00.000  & 7419 - 9067 &  DXe & \dots &  90.0 &  71\\
 06:16:12.9 & 02:00.000  & 7419 - 9067 &  DXo & \dots &  90.0 &  61\\
 06:16:12.9 & 02:00.000  & 7419 - 9067 &  SXe & \dots &  90.0 &  53\\
 06:16:12.9 & 02:00.000  & 7419 - 9067 &  SXo & \dots &  90.0 &  58\\
 06:19:31.0 & 01:56.294  & 7419 - 9067 &  DXe & \dots &  45.0 &  88\\
 06:19:31.0 & 01:56.294  & 7419 - 9067 &  DXo & \dots &  45.0 &  76\\
 06:19:31.0 & 01:56.294  & 7419 - 9067 &  SXe & \dots &  45.0 &  65\\
 06:19:31.0 & 01:56.294  & 7419 - 9067 &  SXo & \dots &  45.0 &  72\\
 06:22:48.9 & 02:38.106  & 7419 - 9067 &  DXe & \dots & 135.0 & 126\\
 06:22:48.9 & 02:38.106  & 7419 - 9067 &  DXo & \dots & 135.0 & 113\\
 06:22:48.9 & 02:38.106  & 7419 - 9067 &  SXe & \dots & 135.0 &  95\\
 06:22:48.9 & 02:38.106  & 7419 - 9067 &  SXo & \dots & 135.0 & 104\\
 \noalign{\smallskip}
\hline
\end{tabular}
\tablefoot{Retarder angles of 45\degr\ and 135\degr\ are for Stokes V. Foster prism position angles of 0\degr\ and 90\degr\ for Stokes Q, and 45\degr\ and 135\degr\ for Stokes U.}
\end{table*}

\setcounter{table}{0}
\begin{table*}[!tbh]
\caption{(continued)}
\begin{tabular}{lllllll}
\hline\hline\noalign{\smallskip}
 UT start    & $t_{\rm exp}$ & $\Delta\lambda$ & Beam & Retarder     & Foster         & $<S/N>$ \\
 (hh:mm:ss)  & (mm:ss)   & (\AA )    &      & \multicolumn{2}{c}{(degrees)} & \\
\noalign{\smallskip}\hline\noalign{\smallskip}
 06:27:19.0 & 01:51.436  & 7419 - 9067 &  DXe & 45.0 &   0.0 & 129\\
 06:27:19.0 & 01:51.436  & 7419 - 9067 &  DXo & 45.0 &   0.0 & 113\\
 06:27:19.0 & 01:51.436  & 7419 - 9067 &  SXe & 45.0 &   0.0 &  91\\
 06:27:19.0 & 01:51.436  & 7419 - 9067 &  SXo & 45.0 &   0.0 & 102\\
 06:30:23.8 & 00:30.417  & 7419 - 9067 &  DXe &135.0 &   0.0 & 134\\
 06:30:23.8 & 00:30.417  & 7419 - 9067 &  DXo &135.0 &   0.0 & 118\\
 06:30:23.8 & 00:30.417  & 7419 - 9067 &  SXe &135.0 &   0.0 &  95\\
 06:30:23.8 & 00:30.417  & 7419 - 9067 &  SXo &135.0 &   0.0 & 105\\
 06:53:14.2 & 00:10.125  & 7419 - 9067 &  DXe & \dots &   0.0 & 174\\
 06:53:14.2 & 00:10.125  & 7419 - 9067 &  DXo & \dots &   0.0 & 155\\
 06:53:14.2 & 00:10.125  & 7419 - 9067 &  SXe & \dots &   0.0 & 133\\
 06:53:14.2 & 00:10.125  & 7419 - 9067 &  SXo & \dots &   0.0 & 148\\
 06:54:45.7 & 00:08.612  & 7419 - 9067 &  DXe & \dots &  90.0 & 174\\
 06:54:45.7 & 00:08.612  & 7419 - 9067 &  DXo & \dots &  90.0 & 156\\
 06:54:45.7 & 00:08.612  & 7419 - 9067 &  SXe & \dots &  90.0 & 134\\
 06:54:45.7 & 00:08.612  & 7419 - 9067 &  SXo & \dots &  90.0 & 148\\
 06:56:12.6 & 00:07.686  & 7419 - 9067 &  DXe & \dots &  45.0 & 187\\
 06:56:12.6 & 00:07.686  & 7419 - 9067 &  DXo & \dots &  45.0 & 167\\
 06:56:12.6 & 00:07.686  & 7419 - 9067 &  SXe & \dots &  45.0 & 145\\
 06:56:12.6 & 00:07.686  & 7419 - 9067 &  SXo & \dots &  45.0 & 161\\
 06:57:43.4 & 00:05.602  & 7419 - 9067 &  DXe & \dots & 135.0 & 200\\
 06:57:43.4 & 00:05.602  & 7419 - 9067 &  DXo & \dots & 135.0 & 181\\
 06:57:43.4 & 00:05.602  & 7419 - 9067 &  SXe & \dots & 135.0 & 157\\
 06:57:43.4 & 00:05.602  & 7419 - 9067 &  SXo & \dots & 135.0 & 172\\
 06:59:42.0 & 00:08.682  & 7419 - 9067 &  DXe & 45.0 &   0.0 & 184\\
 06:59:42.0 & 00:08.682  & 7419 - 9067 &  DXo & 45.0 &   0.0 & 164\\
 06:59:42.0 & 00:08.682  & 7419 - 9067 &  SXe & 45.0 &   0.0 & 142\\
 06:59:42.0 & 00:08.682  & 7419 - 9067 &  SXo & 45.0 &   0.0 & 157\\
 07:01:05.0 & 00:05.598  & 7419 - 9067 &  DXe &135.0 &   0.0 & 201\\
 07:01:05.0 & 00:05.598  & 7419 - 9067 &  DXo &135.0 &   0.0 & 179\\
 07:01:05.0 & 00:05.598  & 7419 - 9067 &  SXe &135.0 &   0.0 & 157\\
 07:01:05.0 & 00:05.598  & 7419 - 9067 &  SXo &135.0 &   0.0 & 173\\
 07:24:28.1 & 00:03.078  & 7419 - 9067 &  DXe & \dots &   0.0 & 204\\
 07:24:28.1 & 00:03.078  & 7419 - 9067 &  DXo & \dots &   0.0 & 183\\
 07:24:28.1 & 00:03.078  & 7419 - 9067 &  SXe & \dots &   0.0 & 154\\
 07:24:28.1 & 00:03.078  & 7419 - 9067 &  SXo & \dots &   0.0 & 171\\
 07:25:52.8 & 00:02.554  & 7419 - 9067 &  DXe & \dots &  90.0 & 185\\
 07:25:52.8 & 00:02.554  & 7419 - 9067 &  DXo & \dots &  90.0 & 167\\
 07:25:52.8 & 00:02.554  & 7419 - 9067 &  SXe & \dots &  90.0 & 140\\
 07:25:52.8 & 00:02.554  & 7419 - 9067 &  SXo & \dots &  90.0 & 153\\
 07:27:13.3 & 00:03.097  & 7419 - 9067 &  DXe & \dots &  45.0 & 226\\
 07:27:13.3 & 00:03.097  & 7419 - 9067 &  DXo & \dots &  45.0 & 203\\
 07:27:13.3 & 00:03.097  & 7419 - 9067 &  SXe & \dots &  45.0 & 174\\
 07:27:13.3 & 00:03.097  & 7419 - 9067 &  SXo & \dots &  45.0 & 192\\
 07:28:38.0 & 00:16.781  & 7419 - 9067 &  DXe & \dots & 135.0 & 168\\
 07:28:38.0 & 00:16.781  & 7419 - 9067 &  DXo & \dots & 135.0 & 152\\
 07:28:38.0 & 00:16.781  & 7419 - 9067 &  SXe & \dots & 135.0 & 127\\
 07:28:38.0 & 00:16.781  & 7419 - 9067 &  SXo & \dots & 135.0 & 140\\
 07:30:48.0 & 00:05.135  & 7419 - 9067 &  DXe & 45.0 &   0.0 & 224\\
 07:30:48.0 & 00:05.135  & 7419 - 9067 &  DXo & 45.0 &   0.0 & 201\\
 07:30:48.0 & 00:05.135  & 7419 - 9067 &  SXe & 45.0 &   0.0 & 173\\
 07:30:48.0 & 00:05.135  & 7419 - 9067 &  SXo & 45.0 &   0.0 & 192\\
 07:32:07.6 & 00:05.146  & 7419 - 9067 &  DXe &135.0 &   0.0 & 197\\
 07:32:07.6 & 00:05.146  & 7419 - 9067 &  DXo &135.0 &   0.0 & 176\\
 07:32:07.6 & 00:05.146  & 7419 - 9067 &  SXe &135.0 &   0.0 & 149\\
 07:32:07.6 & 00:05.146  & 7419 - 9067 &  SXo &135.0 &   0.0 & 165\\
\noalign{\smallskip}
\hline
\end{tabular}
\end{table*}

\begin{table*}[!tbh]
\caption{Heliocentric radial velocities (RV) from Tycho/Moon.}\label{T2-App}
\begin{tabular}{lllllllll}
\hline\hline\noalign{\smallskip}
 JD mid   & UT start   & exp. time & CD: $\lambda$ & $<S/N>$ & $RV(CCF)$ & $\sigma_{\rm RV(CCF)}$ & $RV(corr)$ & $RV(helio)$ \\
 245+     & (hh:mm:ss) & (mm:ss)   & (\AA )        & (p.pix.)& \multicolumn{4}{c}{(\ms )} \\
\noalign{\smallskip}\hline\noalign{\smallskip}
 2458504.6424468 & 03:24:57.4 & 00:20.000 & 3: 4800 - 5441 &    715 &     8.326 &     0.799  &  956.3 &  -947.9\\
 2458504.6424178 & 03:24:57.4 & 00:15.000 & 6: 7419 - 9067 &    532 &    41.834 &     0.466  &  956.3 &  -914.5\\
 2458504.6440000 & 03:27:06.6 & 00:30.000 & 3: 4800 - 5441 &    730 &   -50.114 &     0.798  &  952.6 & -1002.8\\
 2458504.6439421 & 03:27:06.6 & 00:20.004 & 6: 7419 - 9067 &    485 &   -15.239 &     0.168  &  952.8 &  -968.0\\
 2458504.6452708 & 03:28:56.4 & 00:30.000 & 3: 4800 - 5441 &    792 &  -119.032 &     0.815  &  949.7 & -1068.7\\
 2458504.6452130 & 03:28:56.4 & 00:20.000 & 6: 7419 - 9067 &    545 &   -81.132 &     0.494  &  949.8 & -1030.9\\
 2458504.6466725 & 03:30:57.5 & 00:30.000 & 3: 4800 - 5441 &    713 &  -178.394 &     0.825  &  946.4 & -1124.8\\
 2458504.6466146 & 03:30:57.5 & 00:20.000 & 6: 7419 - 9067 &    503 &  -153.111 &     0.191  &  946.5 & -1099.7\\
 2458504.6484109 & 03:33:27.7 & 00:30.000 & 3: 4800 - 5441 &    599 &  -255.226 &     0.854  &  942.4 & -1197.6\\
 2458504.6483530 & 03:33:27.7 & 00:20.000 & 6: 7419 - 9067 &    450 &  -227.549 &     0.498  &  942.5 & -1170.0\\
 2458504.6495961 & 03:35:10.1 & 00:30.000 & 3: 4800 - 5441 &    565 &  -310.599 &     0.882  &  939.6 & -1250.2\\
 2458504.6495382 & 03:35:10.1 & 00:20.000 & 6: 7419 - 9067 &    420 &  -282.169 &     0.210  &  939.7 & -1221.9\\
 2458504.6534144 & 03:40:25.0 & 01:00.000 & 2: 4265 - 4800 &    393 &  -476.019 &     1.121  &  930.8 & -1406.8\\
 2458504.6532407 & 03:40:25.0 & 00:30.000 & 4: 5441 - 6278 &    581 &  -434.441 &     1.110  &  931.2 & -1365.6\\
 2458504.6557107 & 03:43:13.4 & 02:00.000 & 2: 4265 - 4800 &    473 &  -605.956 &     0.966  &  925.5 & -1531.5\\
 2458504.6551898 & 03:43:13.4 & 00:30.000 & 4: 5441 - 6278 &    519 &  -539.598 &     1.346  &  926.7 & -1466.3\\
 2458504.6580266 & 03:46:33.5 & 02:00.000 & 2: 4265 - 4800 &    373 &  -677.804 &     1.081  &  920.2 & -1598.0\\
 2458504.6580266 & 03:46:33.5 & 02:00.000 & 4: 5441 - 6278 &    821 &  -632.076 &     1.427  &  920.2 & -1552.3\\
 2458504.6607164 & 03:50:25.9 & 02:00.000 & 2: 4265 - 4800 &    151 &  -646.447 &     1.445  &  914.1 & -1560.6\\
 2458504.6607164 & 03:50:25.9 & 02:00.000 & 4: 5441 - 6278 &    494 &  -607.877 &     1.657  &  914.1 & -1522.0\\
 2458504.6643518 & 03:53:40.0 & 06:00.000 & 2: 4265 - 4800 &    144 &  -448.866 &     1.209  &  905.9 & -1354.8\\
 2458504.6643518 & 03:53:40.0 & 06:00.000 & 4: 5441 - 6278 &    445 &  -375.057 &     1.298  &  905.9 & -1281.0\\
 2458504.6715417 & 04:04:01.2 & 06:00.000 & 6: 7419 - 9067 &    212 &  -220.415 &     0.668  &  890.0 & -1110.4\\
 2458504.6715428 & 04:04:01.3 & 06:00.000 & 3: 4800 - 5441 &    183 &  -420.888 &     1.214  &  890.0 & -1310.9\\
 2458504.6766493 & 04:11:22.5 & 06:00.000 & 6: 7419 - 9067 &    129 &  -210.267 &     0.995  &  878.9 & -1089.1\\
 2458504.6766505 & 04:11:22.6 & 06:00.000 & 3: 4800 - 5441 &     86 &  -444.900 &     1.870  &  878.9 & -1323.8\\
 2458504.7055718 & 04:46:01.4 & 20:00.000 & 6: 7419 - 9067 &     61 &   305.197 &     4.747  &  820.4 &  -515.2\\
 2458504.7204120 & 05:07:23.6 & 20:00.000 & 6: 7419 - 9067 &     51 &   648.165 &     6.800  &  794.0 &  -145.9\\
 2458504.7356111 & 05:29:16.8 & 20:00.000 & 6: 7419 - 9067 &    108 &   907.567 &     1.649  &  770.2 &   137.4\\
 2458504.7503530 & 05:50:30.5 & 20:00.000 & 3: 4800 - 5441 &     44 &  1677.793 &     3.253  &  750.4 &   927.4\\
 2458504.7503530 & 05:50:30.5 & 20:00.000 & 6: 7419 - 9067 &    190 &  1064.552 &     0.894  &  750.4 &   314.1\\
 2458504.7596250 & 06:12:51.6 & 02:00.000 & 3: 4800 - 5441 &     28 &  1687.196 &     4.677  &  739.9 &   947.3\\
 2458504.7596250 & 06:12:51.6 & 02:00.000 & 6: 7419 - 9067 &     67 &  1349.117 &     1.903  &  739.9 &   609.3\\
 2458504.7619549 & 06:16:12.9 & 02:00.000 & 3: 4800 - 5441 &     37 &  1689.499 &     3.489  &  737.4 &   952.1\\
 2458504.7619549 & 06:16:12.9 & 02:00.000 & 6: 7419 - 9067 &     79 &  1383.665 &     1.391  &  737.4 &   646.2\\
 2458504.7642262 & 06:19:31.0 & 01:56.293 & 3: 4800 - 5441 &     73 &  1707.140 &     2.123  &  735.2 &   972.0\\
 2458504.7642262 & 06:19:31.0 & 01:56.294 & 6: 7419 - 9067 &     98 &  1514.222 &     1.797  &  735.2 &   779.0\\
 2458504.7667587 & 06:22:48.9 & 02:38.106 & 6: 7419 - 9067 &    142 &  1550.885 &     1.121  &  732.8 &   818.1\\
 2458504.7667597 & 06:22:49.0 & 02:38.081 & 3: 4800 - 5441 &    138 &  1681.089 &     1.398  &  732.8 &   948.3\\
 2458504.7696147 & 06:27:19.0 & 01:51.425 & 3: 4800 - 5441 &    137 &  1878.927 &     1.600  &  730.2 &  1148.7\\
 2458504.7696148 & 06:27:19.0 & 01:51.436 & 6: 7419 - 9067 &    146 &  1873.469 &     0.969  &  730.2 &  1143.3\\
 2458504.7712848 & 06:30:23.8 & 00:30.417 & 6: 7419 - 9067 &    150 &  2074.364 &     1.755  &  728.8 &  1345.6\\
 2458504.7712858 & 06:30:23.9 & 00:30.379 & 3: 4800 - 5441 &    138 &  2135.401 &     1.839  &  728.8 &  1406.6\\
 2458504.7729145 & 06:32:50.2 & 00:19.220 & 2: 4265 - 4800 &     45 &  1958.183 &     2.805  &  727.4 &  1230.8\\
 2458504.7729145 & 06:32:50.2 & 00:19.222 & 4: 5441 - 6278 &    169 &  1971.884 &     1.944  &  727.4 &  1244.5\\
 2458504.7740374 & 06:34:31.0 & 00:11.664 & 2: 4265 - 4800 &     57 &  1992.018 &     2.296  &  726.5 &  1265.5\\
 2458504.7740375 & 06:34:31.0 & 00:11.680 & 4: 5441 - 6278 &    185 &  2013.652 &     1.881  &  726.5 &  1287.1\\
 2458504.7750462 & 06:36:00.2 & 00:07.594 & 2: 4265 - 4800 &     57 &  1987.926 &     2.155  &  725.7 &  1262.2\\
 2458504.7750463 & 06:36:00.2 & 00:07.599 & 4: 5441 - 6278 &    176 &  1981.859 &     1.832  &  725.7 &  1256.1\\
 2458504.7760750 & 06:37:29.8 & 00:06.157 & 2: 4265 - 4800 &     66 &  1905.083 &     1.880  &  724.9 &  1180.1\\
 2458504.7760750 & 06:37:29.8 & 00:06.164 & 4: 5441 - 6278 &    173 &  1919.302 &     1.703  &  724.9 &  1194.4\\
 2458504.7775062 & 06:39:28.7 & 00:15.676 & 2: 4265 - 4800 &     84 &  1800.881 &     1.564  &  723.9 &  1077.0\\
 2458504.7775062 & 06:39:28.7 & 00:15.677 & 4: 5441 - 6278 &    299 &  1809.366 &     1.385  &  723.9 &  1085.5\\
 2458504.7785115 & 06:40:58.6 & 00:09.583 & 2: 4265 - 4800 &    102 &  1792.001 &     1.328  &  723.2 &  1068.8\\
 2458504.7785115 & 06:40:58.6 & 00:09.581 & 4: 5441 - 6278 &    327 &  1801.200 &     1.284  &  723.2 &  1078.0\\
 2458504.7799070 & 06:43:00.9 & 00:06.126 & 3: 4800 - 5441 &    238 &  1741.643 &     1.014  &  722.2 &  1019.4\\
 2458504.7799070 & 06:43:00.9 & 00:06.128 & 5: 6278 - 7419 &    391 &  1747.330 &     1.808  &  722.2 &  1025.1\\
 2458504.7809251 & 06:44:28.6 & 00:06.652 & 3: 4800 - 5441 &    235 &  1669.595 &     0.975  &  721.5 &   948.1\\
 2458504.7809251 & 06:44:28.6 & 00:06.660 & 5: 6278 - 7419 &    382 &  1693.253 &     4.371  &  721.5 &   971.7\\
 2458504.7819558 & 06:45:53.3 & 00:15.366 & 3: 4800 - 5441 &    243 &  1573.406 &     1.016  &  720.8 &   852.6\\
 2458504.7819553 & 06:45:53.3 & 00:15.267 & 5: 6278 - 7419 &    380 &  1639.081 &     3.846  &  720.8 &   918.2\\
 2458504.7830391 & 06:47:30.0 & 00:09.164 & 3: 4800 - 5441 &    232 &  1535.844 &     0.993  &  720.2 &   815.7\\
 2458504.7830391 & 06:47:30.0 & 00:09.164 & 5: 6278 - 7419 &    371 &  1621.858 &     4.965  &  720.2 &   901.7\\
 \noalign{\smallskip}
\hline
\end{tabular}
\tablefoot{$RV(helio)=RV(CCF)-RV(corr)$; $\sigma_{\rm RV(CCF)}$ is the standard deviation from the cross-correlation fit (CCF); $RV(corr)$ is the heliocentric and diurnal correction from JPL/Horizons. }
\end{table*}

\setcounter{table}{1}
\begin{table*}[!tbh]
\caption{(continued).}
\begin{tabular}{lllllllll}
\hline\hline\noalign{\smallskip}
 JD mid   & UT start   & exp. time & CD: $\lambda$ & $<S/N>$ & $RV(CCF)$ & $\sigma_{\rm RV(CCF)}$ & $RV(corr)$ & $RV(helio)$ \\
 245+     & (hh:mm:ss) & (mm:ss)   & (\AA )        & (p.pix.)& \multicolumn{4}{c}{(\ms )} \\
\noalign{\smallskip}\hline\noalign{\smallskip}
 2458504.7844493 & 06:49:32.1 & 00:08.646 & 3: 4800 - 5441 &    236 &  1469.923 &     0.974  &  719.3 &   750.6\\
 2458504.7844493 & 06:49:32.1 & 00:08.646 & 5: 6278 - 7419 &    399 &  1520.151 &     1.702  &  719.3 &   800.8\\
 2458504.7854332 & 06:50:54.6 & 00:13.664 & 5: 6278 - 7419 &    408 &  1491.404 &     2.709  &  718.7 &   772.7\\
 2458504.7854350 & 06:50:54.7 & 00:13.778 & 3: 4800 - 5441 &    246 &  1412.585 &     0.945  &  718.7 &   693.8\\
 2458504.7870258 & 06:53:14.2 & 00:09.666 & 3: 4800 - 5441 &    232 &  1336.389 &     0.957  &  717.9 &   618.5\\
 2458504.7870285 & 06:53:14.2 & 00:10.125 & 6: 7419 - 9067 &    197 &  1360.946 &     0.439  &  717.9 &   643.1\\
 2458504.7880788 & 06:54:45.7 & 00:08.612 & 3: 4800 - 5441 &    239 &  1290.583 &     0.941  &  717.3 &   573.3\\
 2458504.7880788 & 06:54:45.7 & 00:08.612 & 6: 7419 - 9067 &    197 &  1329.233 &     0.436  &  717.3 &   611.9\\
 2458504.7890792 & 06:56:12.6 & 00:07.686 & 6: 7419 - 9067 &    214 &  1301.441 &     0.518  &  716.8 &   584.6\\
 2458504.7890801 & 06:56:12.7 & 00:07.642 & 3: 4800 - 5441 &    258 &  1251.405 &     0.913  &  716.8 &   534.6\\
 2458504.7901181 & 06:57:43.4 & 00:05.602 & 6: 7419 - 9067 &    229 &  1247.335 &     0.514  &  716.3 &   531.0\\
 2458504.7901191 & 06:57:43.5 & 00:05.580 & 3: 4800 - 5441 &    274 &  1219.719 &     0.921  &  716.3 &   503.4\\
 2458504.7915085 & 06:59:42.0 & 00:08.668 & 3: 4800 - 5441 &    241 &  1154.335 &     0.963  &  715.7 &   438.7\\
 2458504.7915086 & 06:59:42.0 & 00:08.682 & 6: 7419 - 9067 &    210 &  1165.565 &     0.465  &  715.7 &   449.9\\
 2458504.7924513 & 07:01:05.0 & 00:05.577 & 3: 4800 - 5441 &    259 &  1124.567 &     0.949  &  715.3 &   409.3\\
 2458504.7924514 & 07:01:05.0 & 00:05.598 & 6: 7419 - 9067 &    231 &  1130.427 &     0.521  &  715.3 &   415.2\\
 2458504.7938354 & 07:03:05.6 & 00:03.565 & 2: 4265 - 4800 &    155 &  1036.848 &     1.084  &  714.7 &   322.2\\
 2458504.7938355 & 07:03:05.6 & 00:03.571 & 4: 5441 - 6278 &    340 &  1064.830 &     1.112  &  714.7 &   350.1\\
 2458504.7948287 & 07:04:31.4 & 00:03.590 & 2: 4265 - 4800 &    161 &   999.940 &     0.997  &  714.3 &   285.6\\
 2458504.7948285 & 07:04:31.4 & 00:03.562 & 4: 5441 - 6278 &    350 &  1029.786 &     1.103  &  714.3 &   315.5\\
 2458504.7957700 & 07:05:53.0 & 00:03.051 & 2: 4265 - 4800 &    152 &   967.419 &     1.057  &  714.0 &   253.5\\
 2458504.7957699 & 07:05:53.0 & 00:03.046 & 4: 5441 - 6278 &    333 &  1000.736 &     1.122  &  714.0 &   286.8\\
 2458504.7967590 & 07:07:17.7 & 00:04.552 & 2: 4265 - 4800 &    180 &   933.660 &     1.007  &  713.6 &   220.1\\
 2458504.7967590 & 07:07:17.7 & 00:04.551 & 4: 5441 - 6278 &    388 &   963.221 &     1.105  &  713.6 &   249.6\\
 2458504.7981364 & 07:09:15.9 & 00:06.162 & 2: 4265 - 4800 &    102 &   892.677 &     1.370  &  713.2 &   179.5\\
 2458504.7981364 & 07:09:15.9 & 00:06.163 & 4: 5441 - 6278 &    321 &   922.630 &     1.198  &  713.2 &   209.5\\
 2458504.7990895 & 07:10:36.5 & 00:09.667 & 2: 4265 - 4800 &     94 &   867.117 &     1.504  &  712.9 &   154.2\\
 2458504.7990895 & 07:10:36.5 & 00:09.669 & 4: 5441 - 6278 &    306 &   893.402 &     1.253  &  712.9 &   180.5\\
 2458504.8004906 & 07:12:39.6 & 00:05.578 & 3: 4800 - 5441 &    266 &   851.322 &     1.064  &  712.5 &   138.8\\
 2458504.8004906 & 07:12:39.6 & 00:05.578 & 5: 6278 - 7419 &    414 &   907.438 &     2.912  &  712.5 &   194.9\\
 2458504.8014934 & 07:14:07.0 & 00:04.060 & 3: 4800 - 5441 &    315 &   826.028 &     1.063  &  712.2 &   113.8\\
 2458504.8014933 & 07:14:07.0 & 00:04.043 & 5: 6278 - 7419 &    491 &   906.321 &     1.361  &  712.2 &   194.1\\
 2458504.8024525 & 07:15:30.1 & 00:03.592 & 3: 4800 - 5441 &    291 &   805.790 &     1.092  &  712.0 &    93.8\\
 2458504.8024525 & 07:15:30.1 & 00:03.592 & 5: 6278 - 7419 &    458 &   902.424 &     2.919  &  712.0 &   190.4\\
 2458504.8034412 & 07:16:55.3 & 00:04.042 & 3: 4800 - 5441 &    344 &   787.478 &     1.080  &  711.8 &    75.7\\
 2458504.8034413 & 07:16:55.3 & 00:04.053 & 5: 6278 - 7419 &    538 &   874.264 &     1.047  &  711.8 &   162.4\\
 2458504.8047879 & 07:18:52.4 & 00:02.542 & 3: 4800 - 5441 &    281 &   766.421 &     1.114  &  711.6 &    54.8\\
 2458504.8047879 & 07:18:52.4 & 00:02.545 & 5: 6278 - 7419 &    476 &   876.112 &     1.518  &  711.6 &   164.5\\
 2458504.8056821 & 07:20:08.9 & 00:04.064 & 3: 4800 - 5441 &    334 &   757.858 &     1.095  &  711.4 &    46.4\\
 2458504.8056821 & 07:20:08.9 & 00:04.062 & 5: 6278 - 7419 &    557 &   785.971 &     4.228  &  711.4 &    74.5\\
 2458504.8086762 & 07:24:28.1 & 00:03.046 & 3: 4800 - 5441 &    275 &   759.126 &     1.125  &  711.1 &    48.0\\
 2458504.8086764 & 07:24:28.1 & 00:03.078 & 6: 7419 - 9067 &    235 &   780.699 &     0.445  &  711.1 &    69.6\\
 2458504.8096536 & 07:25:52.8 & 00:02.543 & 3: 4800 - 5441 &    246 &   759.321 &     1.160  &  711.0 &    48.3\\
 2458504.8096537 & 07:25:52.8 & 00:02.554 & 6: 7419 - 9067 &    211 &   765.921 &     0.710  &  711.0 &    54.9\\
 2458504.8105883 & 07:27:13.3 & 00:03.049 & 3: 4800 - 5441 &    308 &   758.627 &     1.114  &  711.0 &    47.7\\
 2458504.8105885 & 07:27:13.3 & 00:03.097 & 6: 7419 - 9067 &    262 &   777.404 &     0.590  &  711.0 &    66.4\\
 2458504.8116480 & 07:28:38.0 & 00:16.781 & 6: 7419 - 9067 &    191 &   795.537 &     0.854  &  710.9 &    84.6\\
 2458504.8116491 & 07:28:38.1 & 00:16.767 & 3: 4800 - 5441 &    245 &   768.041 &     1.157  &  710.9 &    57.1\\
 2458504.8130848 & 07:30:48.0 & 00:05.060 & 3: 4800 - 5441 &    294 &   765.276 &     1.102  &  710.9 &    54.3\\
 2458504.8130853 & 07:30:48.0 & 00:05.135 & 6: 7419 - 9067 &    261 &   791.646 &     0.337  &  710.9 &    80.7\\
 2458504.8140037 & 07:32:07.6 & 00:04.639 & 3: 4800 - 5441 &    240 &   768.576 &     1.156  &  711.0 &    57.6\\
 2458504.8140066 & 07:32:07.6 & 00:05.146 & 6: 7419 - 9067 &    225 &   784.828 &     0.720  &  711.0 &    73.9\\
 2458504.8154003 & 07:34:08.3 & 00:04.571 & 2: 4265 - 4800 &    122 &   752.549 &     1.621  &  711.0 &    41.5\\
 2458504.8154003 & 07:34:08.3 & 00:04.566 & 4: 5441 - 6278 &    303 &   779.056 &     1.347  &  711.0 &    68.0\\
 2458504.8164005 & 07:35:34.2 & 00:05.608 & 2: 4265 - 4800 &     71 &   747.224 &     2.166  &  711.1 &    36.1\\
 2458504.8164006 & 07:35:34.2 & 00:05.624 & 4: 5441 - 6278 &    263 &   771.990 &     1.412  &  711.1 &    60.9\\
 2458504.8173648 & 07:36:58.3 & 00:04.039 & 4: 5441 - 6278 &    296 &   779.078 &     1.387  &  711.2 &    67.9\\
 2458504.8173659 & 07:36:58.4 & 00:04.024 & 2: 4265 - 4800 &    120 &   759.790 &     1.430  &  711.2 &    48.6\\
 2458504.8183656 & 07:38:24.0 & 00:05.583 & 2: 4265 - 4800 &    145 &   762.242 &     1.179  &  711.3 &    50.9\\
 2458504.8183659 & 07:38:24.0 & 00:05.629 & 4: 5441 - 6278 &    315 &   776.904 &     1.358  &  711.3 &    65.6\\
 2458504.8197261 & 07:40:22.3 & 00:04.075 & 2: 4265 - 4800 &     78 &   751.745 &     2.049  &  711.5 &    40.2\\
 2458504.8197262 & 07:40:22.3 & 00:04.077 & 4: 5441 - 6278 &    322 &   780.693 &     1.340  &  711.5 &    69.2\\
 2458504.8206408 & 07:41:41.1 & 00:04.536 & 2: 4265 - 4800 &     75 &   755.874 &     2.147  &  711.7 &    44.2\\
 2458504.8206409 & 07:41:41.1 & 00:04.550 & 4: 5441 - 6278 &    318 &   780.954 &     1.336  &  711.7 &    69.3\\
\noalign{\smallskip}
\hline
\end{tabular}
\end{table*}

\end{document}